\documentclass[apj,twocolumn]{openjournal}
\usepackage{natbib} 
\usepackage[dvipsnames]{xcolor}
\usepackage{aas_macros} 
\usepackage{amssymb}
\usepackage{amsmath}
\usepackage{array}
\usepackage[title]{appendix}
\usepackage{hyperref}	
\hypersetup{colorlinks=true,linkcolor=blue,citecolor=blue,filecolor=blue,urlcolor=blue}
\usepackage[caption=false]{subfig}
\usepackage{soul}                          
\usepackage{xspace}
\usepackage{booktabs}
\usepackage{savesym}
\savesymbol{tablenum}
\usepackage{siunitx}
\usepackage{tabularx}
\restoresymbol{SIX}{tablenum}

\DeclareSIUnit\Msun{\ensuremath{M_\odot}}
\DeclareSIUnit\Zsun{\ensuremath{Z_\odot}}
\DeclareSIUnit\hred{\ensuremath{\textit{h}}}

\newcommand{\eg}{e.g.\ }
\newcommand{\ie}{i.e.\ }

\newcommand{\Eq}[1]{Eq.~(\ref{#1})}

\begin{document}
\title[]{The Impact of Star Formation and Feedback Recipes on the Stellar Mass and Interstellar Medium of High-Redshift Galaxies\vspace{-15mm}}
\author{Harley Katz$^{1,2*}$, Martin P. Rey$^{3}$, Corentin Cadiou$^4$, Taysun Kimm$^5$ \& Oscar Agertz$^4$}
\thanks{$^*$E-mail: \href{mailto:harleykatz@uchicago.edu}{harleykatz@uchicago.edu}}

\affiliation{$^{1}$Department of Astronomy \& Astrophysics, University of Chicago, 5640 S Ellis Avenue, Chicago, IL 60637, USA}
\affiliation{$^{2}$ Kavli Institute for Cosmological Physics, University of Chicago, Chicago IL 60637, USA}
\affiliation{$^{3}$Sub-department of Astrophysics, University of Oxford, Keble Road, Oxford OX1 3RH, United Kingdom}
\affiliation{$^{4}$Lund Observatory, Division of Astrophysics, Department of Physics, Lund University, Box 43, SE-221 00 Lund, Sweden}
\affiliation{$^{5}$Department of Astronomy, Yonsei University, 50 Yonsei-ro, Seodaemun-gu, Seoul 03722, Republic of Korea}

\begin{abstract}
We introduce {\small MEGATRON}, a new galaxy formation model for cosmological radiation hydrodynamics simulations of high-redshift galaxies. The model accounts for the non-equilibrium chemistry and heating/cooling processes of $\geq 80$ atoms, ions, and molecules, coupled to on-the-fly radiation transfer. We apply the model in a cosmological setting to the formation of a $\SI{e9}{\Msun}$ halo at $z=6$, and run 25 realizations at pc-scale resolution, varying numerous parameters associated with our state-of-the-art star formation, stellar feedback, and chemical enrichment models. We show that the overall budget of feedback energy is the key parameter that controls star formation regulation at high redshift, with other numerical parameters (\eg supernova clustering, star formation conditions) having a more limited impact. As a similar feedback model has been shown to produce realistic $z=0$ galaxies, our work demonstrates that calibration at $z=0$ does not guarantee strong regulation of star formation at high-redshift. Interestingly, we find that subgrid model variations that have little impact on the final $z=6$ stellar mass can lead to substantial changes on the observable properties of high-redshift galaxies. For example, different star formation models based on, e.g.\ density thresholds or turbulence inspired criteria, lead to fundamentally distinct nebular emission line ratios across the interstellar medium (ISM). These results highlight the ISM as an important resource for constraining models of star formation, feedback, and galaxy formation in the JWST era, where emission line measurements for $>1,000$ high-redshift galaxies are now available.
\end{abstract}
\keywords{high-redshift galaxies, ISM, galaxy formation}

\section{Introduction}
The launch of JWST \citep{Gardner2006,Gardner2023} has ushered in a new era for high-redshift galaxy formation. The spectroscopic redshift barrier set by HST at $z\sim11$ \citep{Oesch2016} has been far surpassed \citep[e.g.][]{Curtis-Lake2023,BWang2023,Carniani2024}, while emission lines that provide detailed insight into the ISM have been detected to $z>12$ \citep{Zavala2024,Castellano2024}. 

This new data immediately challenged conventional understanding of galaxy formation in the early Universe. For example, numerous high-redshift galaxy surveys with JWST have reported the detection of an overabundance of bright galaxies at $z\gtrsim9$ compared to earlier constraints from Hubble and most theoretical models \citep[e.g.][although c.f. \citealt{Willott2023}]{Finkelstein2023_b,Harikane2024,Leung2023,Chemerynska2023}. Likewise, massive quiescent galaxies at $z\sim3-5$ appear to form at exceptionally early epochs \citep[e.g.][]{deGraaff2024,Glazebrook2024}. Many early galaxies exhibit chemical abundance patterns that have never been seen in the gas-phase \citep[e.g.][]{Cameron2023_nitrogen,Senchyna2024,Isobe2023_nitrogen,Topping2024}, which may indicate exotic enrichment channels, e.g.\ from supermassive stars \citep[e.g.][]{Charbonnel2023,Chaves2024}. Moreover the ISM gas temperatures in early galaxies may be at odds with standard heating mechanisms \citep[e.g.][]{Katz2023_jwst,Laseter2024}. Finally, some of the high-redshift JWST spectra appear to require exotic stellar populations or a top-heavy stellar initial mass function \citep[e.g.][]{Cameron2023_NDG,Katz2024}.

These new anomalies supplement a previous set of inconsistencies between simulations and observations that arose from HST/ALMA data including the presence of high-redshift Balmer breaks \citep{Hashimoto2018}, extremely high [O~{\small III}]~88~$\mu$m/[C~{\small II}]~158~$\mu$m ratios \citep{Harikane2020}, or very extended cold gas haloes \citep{Fujimoto2020}. 

Understanding the physical mechanisms responsible for the discrepancies between simulations and observations remains a key theoretical challenge. For example, numerous solutions to the overabundance of bright galaxies have been proposed including changes to the star formation efficiency \citep[e.g.][]{Dekel2023,Feldmann2025}, the absence of dust \citep{Ferrara2023}, $M_{\rm UV}$ variability \citep[e.g.][]{Mason2023,Shen2023,Sun2023,Kravtsov2024}, IMF variations \citep[e.g.][]{Yung2024}, chemical anomalies \citep{Katz2024He}, or nebular emission \citep{Katz2024}.

Prior to JWST, numerous numerical simulations of the early Universe such as {\small AURORA} \citep{Pawlik2017}, {\small CoDa} \citep{Ocvirk2020}, {\small CROC} \citep{Gnedin2014}, {\small FIRE} \citep{Ma2018}, {\small FIREbox HR} \citep{Feldmann2025}, {\small FirstLight} \citep{Ceverino2017}, \citep{Ma2018}, {\small FLARES} \citep{Lovell2021}, {\small OBELISK} \citep{Trebitsch2021}, {\small RENAISSANCE} \citep{Oshea2015}, \small{SERRA} \citep{Pallottini2022}, {\small SPHINX} \citep{Rosdahl2018}, and {\small THESAN} \citep{Kannan2022} were very successful in reproducing observational data such as the UV luminosity function, despite fundamental differences in resolution (spatial and mass), subgrid models for star formation, feedback, and metal enrichment, and included physics such as radiative transfer, non-equilibrium cooling, and dust. This demonstrates that the constraining power of the UV luminosity function is rather limited. 

Part of the reason why UV luminosity has poor constraining power is that it is generally only sensitive to the younger stellar populations in the galaxy \citep[e.g.][]{Kennicutt2012} rather than the full star formation history. For this reason, the inferred stellar mass-halo mass relation \citep[e.g.][]{Moster2018,Behroozi2019} has been a common benchmark. However, neither stellar mass nor halo mass are directly observable quantities, rather they are inferred from observations under a set of assumptions that may differ from those used in models. This can lead to systematic errors in comparing inferred and modeled stellar mass-halo mass relations. Furthermore, no pre-JWST simulation reliably predicted the stellar mass function in the interval $z\sim4-9$ \citep{Weibel2024}. Stellar mass in particular has order of magnitude uncertainties depending on choice of prior of star formation history \citep{Narayanan2024} and this manifests in studies that observe galaxies with similar redshifts and fluxes predicting vastly different stellar masses \citep{spdrv1}. Convergence also has not yet been reached in studies comparing spatially resolved and integrated stellar masses \citep{GA2023,GA2024}. Most importantly, the models used to infer the stellar mass-halo mass relation are inconsistent with the new high-redshift JWST data \citep{Finkelstein2023_b}. Because JWST can see much redder than HST, rather than using inferred stellar masses, constraints on the older stellar populations are available via photometry (and/or spectra), if they can be forward modeled.  

In addition to observing older stellar populations at longer wavelengths, another advantage of JWST is the ability to probe into the ISM of early galaxies for large samples \citep[e.g.][]{Cameron2023,Sanders2023}, which complements what can be done with ALMA for much fewer numbers of galaxies \citep[e.g.][]{Carniani2025_alma,Schouws2025}. These studies have demonstrated that the nebular conditions in the early Universe are vastly different compared to those in local galaxies. Comparing simulations with the observed ISM conditions at high redshift via rest-frame UV and optical emission lines represents a new opportunity to test the physics present in numerical simulations.

However, among the larger-scale simulation suites, only a few projects such as {\small FIRE}, {\small FIREBox HR}, {\small Obelisk}, {\small SERRA}, {\small SPHINX}, and {\small RENAISSANCE} have the resolution to begin to probe the ISM (at the cost of modeling either a smaller volume, fewer galaxies, or stopping the simulation at high-redshift). Such simulations often use more simplistic models than what is considered the state-of-the-art in the ISM simulation community \citep[]{Kim2023}. Simulations with detailed ISM models are often limited to either isolated or small high-redshift dwarf galaxies \citep[e.g.][]{SS2014,Hu2016,Steinwandel2020,Hislop2022,PRISM,Deng2024}, which limits the possible comparison between the models and JWST observations. Moreover, much of the detailed information from JWST comes in the form of emission lines from both primordial species (\eg H$\alpha$, H$\beta$, He~{\small II}~$\lambda$1640) and metals (\eg [O~{\small III}]~$\lambda$5007, [O~{\small II}]~$\lambda\lambda$3727, [C~{\small III}]~$\lambda\lambda$1908), many of which are not self-consistently predicted by the current generation of simulations. 

Predicting emission lines (and more generally the full spectra of galaxies) requires knowing the detailed properties of the radiation field and thermodynamics of both ionized and neutral gas. Cosmological radiation hydrodynamics codes often include non-equilibrium chemistry for primordial species \citep[e.g.][]{Pawlik2008,Rosdahl2013,Gnedin2014}, but few include non-equilibrium metal chemistry, the effects of which have recently been shown to be important for some emission lines \citep[e.g.][]{Richings2022,PRISM}. Resolving the detailed structure of the H~{\small II} regions is also crucial \citep[e.g.][]{Jin2022} and this remains well beyond the capabilities of current simulations of full galaxies. Nevertheless, in order to take advantage of the most recent JWST data as a means of both understanding the physics that governs our own Universe and for constraining theoretical models of galaxy formation, detailed galaxy-scale models of the ISM are required to forward-model the observations.  

For this reason, in this work we adopt the {\small RAMSES-RTZ} code \citep{RTZ} to distill the physics from numerics of early galaxy formation. {\small RAMSES-RTZ} is unique among galaxy formation codes as it not only solves for gravity, hydrodynamics, radiation transfer, and primordial chemistry, but it can also follow the non-equilibrium chemistry of metals, fully coupled to the radiative transfer, which is the key component for predicting emission lines. The large chemical network drastically increases the number of emission lines that can be predicted in a non-equilibrium manner compared to previous generations of simulations \citep[e.g.][]{Lupi2020}. We develop a new, state-of-the-art galaxy formation model and systematically test how modifications to sub-grid parameters, feedback physics, and resolution, impact the stellar mass growth and the ISM of a single high-redshift dwarf galaxy. We demonstrate how the ISM, in particular observable emission line ratios, exhibit strong sensitivity to subtle changes in subgrid recipes and feedback, and thus represent a new means of constraining models of galaxy formation.

\section{Methods}
In this work, we describe a new model for high-redshift galaxy formation: {\small MEGATRON}. The model is developed as part of the {\small RAMSES-RTZ} code \citep{RTZ}, based on {\small RAMSES} \citep{Teyssier2002} and {\small RAMSES-RT} \citep{Rosdahl2013,Rosdahl2015}. In this section, we outline our fiducial model and a broad parameter study of subgrid physics and numerical resolution.
 
\subsection{Gravity \& Hydrodynamics}
We numerically solve the equation of hydrodynamics in the simulation with a closure between gas pressure and internal energy with an adiabatic index of $\gamma = 5/3$, suited for an ideal monoatomic gas. The associated non-linear conservation laws are evolved with a MUSCL-Hancock scheme \citep{vannoordvanvanleerUltimateConservativeDifference1979} using a linear reconstruction of the conservative variables at cell interfaces with a MinMod total variation diminishing scheme, and using the HLLC approximate Riemann solver \citep{toro_RiemannSolversNumerical_2009}. We use a Courant factor of $0.8$. Dynamics of collisionless star and dark matter (DM) particles are computed using a particle-mesh solver and we compute their contribution to the gravitational potential using cloud-in-cell interpolation. Gravitational accelerations are calculated using a finite-difference of the potential; the latter is computed from the total density $\rho = \rho_\mathrm{gas} + \rho_\star + \rho_\mathrm{DM}$ using a multigrid scheme \citep{guillet_SimpleMultigridScheme_2011}. 

\subsection{Chemistry \& Thermodynamics}
Our model is unique among large cosmological simulations in that it employs the {\small PRISM} model \citep{PRISM}, which includes a large non-equilibrium thermochemistry network for primordial species, metals, and molecules that are fully coupled to on-the-fly multi-frequency radiation transfer. Compared to previous studies, improved modeling of the gas thermochemistry allows us to more accurately predict the non-equilibrium emission and absorption from the ISM and CGM. 

\subsubsection{Chemical Model}
In total we consider 82 species for our non-equilibrium calculation:
\begin{itemize}
    \item \underline{Primordial species}: H~{\small I}-{\small II}, He~{\small I}-{\small III}, $e^{-}$,
    \item \underline{Metal ions}: C~{\small I}-{\small VI}, N~{\small I}-{\small VII}, O~{\small I}-{\small VIII}, Ne~{\small I}-{\small X}, Mg~{\small I}-{\small X}, Si~{\small I}-{\small XI}, S~{\small I}-{\small XI}, \& Fe~{\small I}-{\small XI},
    \item \underline{Molecules}: H$_2$ \& CO.
\end{itemize}
While {\small RAMSES-RTZ} is capable of handling all ionization states of each metal ion, we have restricted ourselves to what we define as the ``minimal cooling model''. Using {\small CLOUDY} \citep{Ferland2017}, we calculate the contribution of each metal ion to the total equilibrium cooling rate in the temperature range $10-\SI{e9}{K}$ under the assumption of a \cite{Haardt2012} UV background at $z=2$. We only model in non-equilibrium those ions that contribute at least \SI{1}{\percent} to the total equilibrium cooling rate across the aforementioned temperature range. This decreases by \SI{40}{\percent} the number of metal ions to be followed out-of-equilibrium, easing the memory requirements to adapt our model to modern supercomputers. All ions that are not followed in non-equilibrium are still considered under the assumption of collisional ionization equilibrium. 

The non-equilibrium evolution of the aforementioned species are modelled as described in \cite{RTZ,PRISM}. More specifically, we account for ionization, recombination, and charge exchange processes. For H and He, collisional ionization and radiative and dielectronic recombination are adopted following \cite{Rosdahl2013} based on rates from \cite{Cen1992} and \cite{Hui1997}, respectively. Radiative and dielectronic recombination recombination rates for metals are largely taken from \cite{Badnell2006} and \cite{Badnell2003}, respectively, with the exception of certain higher ionization states\footnote{We rely on \cite{Badnell2003,Badnell2006} where possible. If the species is not available in those datasets, we revert to older data.} of heavy elements where we rely on \cite{Aldrovandi1973,Shull1982,Arnaud1985,Arnaud1992}. Collisional ionization rate fitting functions for metals are adopted from \cite{Voronov1997}. Charge exchange reaction rates are compiled from \cite{Kingdon1996,Stancil1999,Barragan2006}. Recombination on the surface of dust grains is computed using reaction rates taken from \cite{Weingartner2001}. All atoms and their ions are subject to photoionization. Cross sections for all atomic species and their ions are calculated using the analytic fits from \cite{Verner1996}. Although the {\small PRISM} model can also handle ionization from cosmic-rays, for this work, we neglect their effects, as they will be considered in future work using a fully-coupled CR solver \citep[e.g.][]{RM2024}.  

The formation and destruction of H$_2$ and CO molecules is described in \cite{Katz2017,PRISM}. H$_2$ formation proceeds via two channels: primordial reactions stimulated by H$^-$ and formation on dust. Primordial rates follow those presented in \cite{Glover2010} while dust formation rates are a combination of \cite{Gnedin2009} and \cite{Bialy2019} (see \citealt{PRISM}). Due to finite spatial resolution, we artificially enhance the formation rate of H$_2$ by a factor of 10 (see also \citealt{Gnedin2009}) to account for unresolved, sub-grid clumping. Note that such small-scale clumping is expected to occur below the grid scale (which is a $\sim1-20$~pc for the simulations presented here) due to unresolved turbulence that typically leads to a log-normal gas density probability distribution function. Destruction of H$_2$ can occur through collisions or via photodissociation (direct and via recombinative dissociation). Collisional dissociation rates are adopted from \cite{Glover2008} while cross-sections in the Lyman-Werner and ionizing bands are as described in \cite{Baczynski2015}. We include H$_2$ self-shielding at outlined \cite{Katz2017}.

CO formation and destruction is based on \cite{Nelson1997,Glover2012} with modifications as described in \cite{PRISM}. The physical mechanism of CO formation is assumed to be initiated by H$_2$ and C$^+$ which goes on to form CH$_x$ molecules. The rate limiting steps are then dictated by the survivability of CH$_x$ molecules. Destruction of CO and CH$_x$ molecules are assumed to be dominated by far UV radiation with destruction rates adopted from \cite{Heays2017}. We note that, in principle, the formation rate of CO is subject to the same sub-grid clumping as assumed for H$_2$. However, since H$_2$ is the primary reactant in the formation of CO, our subgrid clumping factor also propagates into the formation of CO. In previous work, we focused on the ISM and thus neglected collisional destruction of CO since photodissociation by UV radiation is the dominant process in such environments. In this work, we supplemented the model to account for collisional destruction of CO by electrons in high-temperature gas. This provides a destruction path for CO ejected from the galaxy in a wind, where the radiation field is weak and the temperature is high\footnote{More specifically, we adopt the UMIST rate \citep{McElroy2013} for CO$^+$~+~$e^-\rightarrow$~C~+~O. This rate naively over-predicts the destruction mechanism that we are interested in due to the additional Coulomb attraction and thus we exponentially suppress the rate below temperatures of \SI{2e5}{K}.}.

Much of the chemistry described above relies on knowing the dust content in each gas cell in the simulation. We employ an empirical model where the dust-to-gas mass ratio is assumed to follow a broken power-law as a function of metallicity following observations of the local Universe \citep{RR2014}. To remain consistent with these observations, the composition of the dust is that of the BARE-GR-S model of \cite{Zubko2004}. We assume that dust is only present in cells with $T<\SI{e6}{K}$ and use the chemical composition of the BARE-GR-S dust model to self-consistently compute the depletion factors of each metal into the dust phase. As described in \cite{PRISM}, because elements are enriched individually, metal abundance ratios can significantly deviate from solar abundance patterns. To avoid unphysical depletion rates, we cap the maximum depletion rate of each metal to those at solar metallicity defined in \cite{Dopita2000,Grevesse2010}. Future work will self-consistently couple dust grains to the thermochemistry (Rodriguez-Montero~et~al.~{\it in prep.}).

\subsubsection{Heating \& Cooling}
{\small PRISM} has been designed to reproduce the expected equilibrium of the local ISM \citep[e.g.][]{Wolfire1995,Wolfire2003} but generalized into a 3D non-equilibrium galaxy formation setting (see Figure~1 of \citealt{PRISM}).
Our approach is here fundamentally different from other large simulations, where radiative heating and cooling is often determined by combining a UV background and {\small CLOUDY} tables \citep[e.g.][]{Schaye2015,Pillepich2018,Dubois2014,Feldmann2023}.
This is motivated by the fact that, in 1D tests, such models often fail to reproduce the equilibrium pressure-density relation observed in the local ISM \citep{Kim2023}. Note that a subset of simulations have attempted to account for either the local radiation field or non-equilibrium effects in the context of cosmological simulations \citep[e.g.][]{Gnedin2012,Oppenheimer2018,Capelo2018,Lupi2020}; however, the models are either specifically designed for the warm and cold ISM, ignore the local radiation field, do not track higher ionization states of metals that matter for the CGM or ionized gas more generally, or do not follow the impact of individual elements on the cooling function.

Our full heating and cooling model is nearly identical to that presented in \cite{PRISM}. It includes photoheating \citep{Rosdahl2013}, photoelectric heating \citep{Bakes1994,Wolfire1995,Wolfire2003} with corrections due to a different assumed PAH abundance \citep{Zubko2004}, and heating from H$_2$ formation \citep{Sternberg1989,Rollig2006} and H$_2$ excitation/dissociation \citep{Black1977,Draine1996,Burton1990}. 

We account for cooling from primordial species \citep{Cen1992,Hui1997,Rosdahl2013,Osterbrock2006,Haiman1996,Black1981}, H$_2$ \citep{Moseley2021}, CO \citep{McKee1982,Hollenbach1989,Koyama2000}, dust recombination \citep{Wolfire2003}, dust-gas collisional processes \citep{Hollenbach1989,Draine2011,Bialy2019}, and metal lines. At $T<\SI{e4}{K}$, cooling from metal lines is calculated for C~{\small I}, C~{\small II},  N~{\small II}, O~{\small I}, O~{\small III}, Ne~{\small II}, Si~{\small I}, Si~{\small II}, S~{\small I}, Fe~{\small I}, and Fe~{\small II}, under the approximation of a two- or three-level ion as appropriate. We consider the following collisional partners: H, H$^+$, ortho and para H$_2$, $e^-$, He, He$^+$, and He$^{++}$, with collision rates adopted from \cite{Ferland2017}. At $T>\SI{e3}{K}$, we have created tables of cooling rates from {\small CLOUDY} \citep{Ferland2017} that are interpolated ion-by-ion assuming that electrons are the only collisional partner. In the regime $10^3\leq T/\si{K}\leq10^4$, we subtract the cooling contribution from the ions that we compute explicitly from the cooling tables to prevent double counting. At $T\sim\SI{e4}{K}$, we smoothly interpolate between the explicit computation and the cooling tables using the method presented in \cite{Bovino2016}. 
We assume collisional ionization equilibrium (CIE) for ions that are followed in equilibrium, and we interpolate our cooling tables to account for their impact on the gas thermochemistry \citep[e.g.][]{Gray2015}. 

All chemical reactions, temperature, and photon group (see Table~\ref{tab:engy_bins}) updates are evolved using a semi-implicit method originally based on \cite{Anninos1997}. We adapt the timestep following a \SI{10}{\percent} rule \citep[e.g.][]{Rosdahl2013} so that no quantity changes by more than \SI{10}{\percent} in each cooling subcycle. This method has been shown to reproduce the behaviour of more complex implicit solvers at a much reduced computational cost \citep{RTZ}.

\begin{figure}
    \centering
    \includegraphics[width=0.45\textwidth]{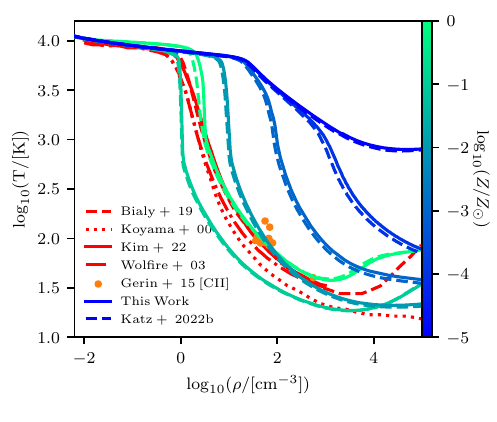}
    \includegraphics[width=0.45\textwidth]{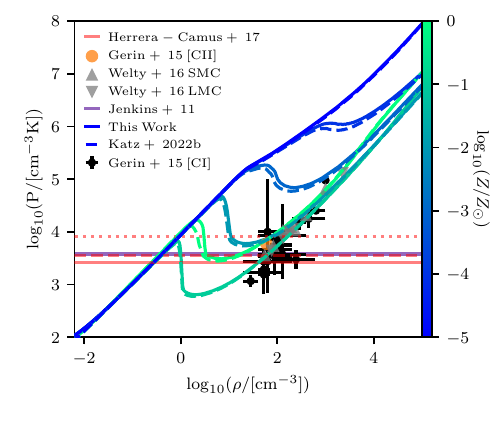}
    \caption{(Top) Equilibrium temperature-density (top) and pressure-density (bottom) relations for the thermodynamic model used in MEGATRON, as a function of metallicity (solid lines). For all models, we assume a local ISM radiation field such that $G_0=1$ and a background cosmic ray hydrogen ionization rate of $10^{-16}\ {\rm s^{-1}}$. At solar metallicities (green), the thermodynamic model is consistent with other ISM models (\protect\cite{Bialy2019}, \protect\cite{Koyama2000}, \protect\cite{Kim2023}; red dashed, dotted, and solid, respectively) and  data in the Milky Way and the LMC/SMC (\protect\cite{Gerin2015}, \protect\citep{Jenkins2011}, \protect\cite{Herrera2017}, \protect\cite{Welty2016}; orange points, purple line, red dashed, solid, dotted lines for cold-neutral fractions of 0.7, 0.5, and 0.3, and grey triangles, respectively). Changes in atomic data and reaction rates lead to small variations with the earlier version of the model (\protect\citealt{PRISM}, dashed blue lines).}
    \label{fig:trho}
\end{figure}

Note that the only major change\footnote{Other minor changes include updating the H$_2$ cooling function from \cite{Hollenbach1979,Galli1998} to \cite{Moseley2021} and fixing a bug in our dust recombination cooling where a factor of $0.5$ was inadvertently missing.} in cooling from \cite{PRISM} is the extension of the cooling tables below \SI{e4}{K}. This change has little impact on the equilibrium pressure-density curve at solar ISM conditions because the cooling is completely dominated by [C~{\small II}] and [O~{\small I}] lines for which the new computation is identical. This can be seen in Figure~\ref{fig:trho} where we show the equilibrium temperature-density (top) and pressure-density (bottom) relations as a function of metallicity comparing the updates applied in this work (solid lines) to that from \cite{PRISM} (dashed lines). Almost no difference is seen between the models and both agree with observational constraints at solar metallicity because this test predominantly probes the physics of neutral gas. 

\begin{figure}
    \centering
    \includegraphics[width=0.45\textwidth]{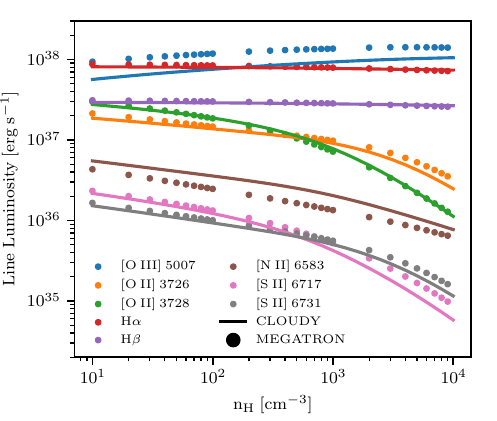}
    \caption{Emission line luminosities from a Strömgren sphere around an O4V star in gas with varying density at \SI{0.1}{\Zsun} metallicity. Results from the MEGATRON non-equilibrium, radiative transfer calculation (points) show good agreement with CLOUDY calculations (lines).}
    \label{fig:line_cal}
\end{figure}

Rather, the extension of the tables results in more accurate cooling in metal-enriched ionized regions where significant cooling contributions from more highly-excited energy levels and ions may be important\footnote{For example, in an example solar-metallicity CLOUDY model where the ionized gas temperature is $\sim8,000$~K in the center of the nebula, ions such as S~IV, S~III, Ne~III, N~III, Ar~III, Fe~III, O~II, S~II, N~II, Mg~II, and others all contribute more than 1\% of the total cooling rate at some radius in the nebula and hence are likely important for the ionized gas temperature. For this reason, our new cooling model now better captures these ions at $T<10^4$~K. Note that depletion onto dust and dust model can significantly impact which ions are important and this is self-consistently computed in the MEGATRON model.}. To demonstrate the accuracy of the updated model, we have run a series of Strömgren sphere tests where we place a single O4V star modelled as having a blackbody temperature of 42,900~K\footnote{Stellar parameters for this star can be found at \url{https://www.pas.rochester.edu/~emamajek/EEM_dwarf_UBVIJHK_colors_Teff.txt}.} \citep{Pecaut2013} at the corner of a uniform density cube of gas initialized at \SI{100}{K} and \SI{0.1}{\Zsun} metallicity and evolve the radiation field and thermochemistry of the gas until the system reaches equilibrium. Note that hydrodynamics and gravity is excluded from this experiment. We then calculate the emission line luminosities for various strong lines and compare the results to a similar 1D simulation with {\small CLOUDY}. In Figure~\ref{fig:line_cal}, we compare the results from our model (points) with those from {\small CLOUDY} (lines). For nearly all lines, the predicted emission line luminosities are in very good agreement. Subtle differences are present for a few reasons. First, we have adopted the default atomic data in {\small PyNeb} \citep{Luridiana2015} which can differ from {\small CLOUDY}. Second, the dust prescription in {\small CLOUDY} is different from our model, which has an impact on temperature. We have specifically adopted \SI{0.1}{\Zsun} for this experiment so that dust has an impact on the results but does not necessarily dominate. This has a noticeable effect on the [O~{\small III}]~$\lambda$5007 line as the gas inside the H~{\small II} region in our model is slightly hotter than in {\small CLOUDY}. Due to the exponential scaling of the emissivity curve for collisionally excited lines, small changes in temperature can lead to significant changes in luminosity. H$\alpha$ and H$\beta$ are much less sensitive to temperature which is why we find such good agreement in the ionized region. Third, we have adopted slightly different values for depletion onto dust in our model than the default in {\small CLOUDY}. This has the strongest impact on the nitrogen emission lines where our adopted depletion value is higher. The magnitude of the difference we observe in our code compared to {\small CLOUDY} is similar to the differences between {\small CLOUDY} and {\small MAPPINGS} \citep{DAgostino2019}\footnote{Across all emission lines in Figure~\ref{fig:line_cal}, we find a median difference of 0.035~dex between our code and CLOUDY with a maximum difference of 0.22~dex. For comparison, \cite{DAgostino2019} found a $~0.1$~dex difference between CLOUDY and MAPPINGS for the BPT diagram.}. Hence, we are confident that our model is well-calibrated in both neutral and ionized regimes.

\subsection{Refinement Strategy}
We use an adaptive mesh refinement strategy to split grid cells and increase the dynamic range of the simulation. In our fiducial model, refinement is allowed down to \SI{76}{pc}, constant in comoving units. In other words, the maximum spatial resolution of the simulation slowly degrades with time, going from \SI{3.6}{pc} at $z=20$ to \SI{11}{pc} at $z=6$. This avoids both numerical effects of releasing grid levels at specified times (\eg \citealt{Snaith2018}), and also affords higher spatial resolution at high-redshifts when haloes are much smaller in mass and radius. We combine a quasi-Lagrangian refinement strategy with additional super-Lagrangian refinement criteria targeting the Jeans' length. A cell is flagged for refinement into 8 equally-sized children cells if
\begin{enumerate}
    \item either the cell mass exceeds $M_{\mathrm{DM,cell}}+\frac{\Omega_\mathrm{m}}{\Omega_\mathrm{b}}M_{\mathrm{b,cell}}>8 \, m_\mathrm{DM}$, where $M_{\mathrm{DM,cell}}$ and $M_{\mathrm{b,cell}}$ are respectively the DM and baryon (stars + gas) mass in the current cell,
    \item or the cell size is at least a quarter of the local Jeans length 
    \begin{equation}
        \lambda_\mathrm{J} = \sqrt{\frac{\pi c_\mathrm{s}^2}{G\rho}},
    \end{equation}
    where $c_\mathrm{s} = \sqrt{\gamma k_\mathrm{b}T/m_\mathrm{p}}$ is the sound speed, $G$ is the gravitational constant, $\rho$ is the gas density and $m_\mathrm{p}$ is the proton mass,
    \item and the value of the passive scalar (that identifies cells in the Lagrange volume) in the cell is larger than 0.9\footnote{When we create the initial conditions of the zoom simulation, all cells that are on the maximum level of refinement are tagged with a passive scalar of 1, while all other cells have this value set to 0. This passive scalar follows gas flows, allowing us to easily identify where the Lagrange region is throughout the simulation.}.
\end{enumerate}
When refining cells, we directly inject the conservative variables (density, momentum, etc.) into the children cells.

We use adaptive time stepping, where the time step is halved wherever resolution is doubled.
In the following, we will refer to the `fine' time step as the shortest time step, and as the `coarse' time step as the time step of the base grid (\SI{39}{kpc} comoving).

\vspace{2mm}
\subsection{Star Formation}
\label{sec:sf_alg}
Once the gas has cooled and condensed sufficiently, we consider several approaches to spawning star particles in the simulation.  

\subsubsection{Turbulence Inspired Star Formation}
Our fiducial algorithm assumes that star formation occurs in the simulation based on estimates of the local turbulent properties of the gas \citep{Kimm2017}. The algorithm is inspired by high-resolution simulations of turbulent star forming gas clouds \citep[e.g.][]{Padoan2011,Federrath2012}. The method is identical to that presented in \cite{Rosdahl2018} and similar to what was used in \cite{Agertz2021}.

We consider a gas cell eligible for star formation if it satisfies the following criteria:
\begin{enumerate}
    \item The gas density is $>\SI{10}{cm^{-3}}$ and $>200\times$ the mean background. The former threshold has negligible impact on the simulation and is set purely for computational efficiency while the latter is designed so that stars do not form outside haloes at unrealistically high redshifts.
    \item The turbulent Jeans length (see Equation~2 of \citealt{Rosdahl2018} or alternatively \citealt{Bonazzola1987,Federrath2012}) is smaller then the cell size (\ie unresolved).
    \item The gas cell is a local density maximum compared to the six neighbouring cells.
    \item The gas flow is locally converging (as measured from the six neighbouring cells). 
\end{enumerate}

When a gas cell is flagged for star formation the star formation rate density is computed as 
\begin{equation}
    \dot{\rho}_*=\epsilon_*\frac{\rho}{t_{\rm ff}},
\end{equation}
where $\rho$ is the gas density of the cell, $t_{\rm ff}$ is the gas free-fall time, and $\epsilon_*$ is the efficiency by which the gas is converted into stars over a free-fall time. The efficiency is estimated again from the local turbulent gas properties such that
\begin{equation}
\epsilon_{\rm *} = \frac{\epsilon_{\rm acc}}{2 \phi_t} \exp\left(\frac{3}{8}\sigma_s^2 \right) \left[ 1 + {\rm erf} \left(\frac{\sigma_s^2 - s_{\rm crit}}{\sqrt{2\sigma_s^2}}\right)\right],
\label{eq:sfe}
\end{equation}
where $\epsilon_{\rm acc}=0.5$ is the accretion fraction of gas onto stars, $1/\phi_t=0.57$, $\sigma_s^2 = \ln \left(1 + b^2 \mathcal{M}^2 \right)$, $b=0.4$ (which assumes a mixture of compressive and solenoidal modes), $\mathcal{M}$ is the Mach number (see also \citealt{Padoan2011,Federrath2012}) which is computed from the local properties of the gas. Finally 
\begin{equation}
s_{\rm crit} = {\rm ln} \left(0.067\theta^{-2}\,\alpha_{\rm vir} \,\mathcal{M}^2\right),
\end{equation}
where $\theta=0.33$ accounts for shock thickness with respect to cloud size \citep{Federrath2012}, and $\alpha_{\rm vir}\equiv 2 E_{\rm kin}/ |E_{\rm grav}|$ is the virial parameter of a cloud.

The number of star particles created in each cell is drawn from a Poisson distribution with a typical value of $N_*=\dot{\rho}_*\Delta t\Delta x^3/M_*$ \citep[e.g.][]{Rasera2006}. Here $\Delta t$ is the time step, $\Delta x$ is the physical size of the gas cell, and $M_*=\SI{500}{\Msun}$ is the mass of an individual star particle. Note that below we will consider variations to this fiducial choice for the particle mass.

In this work, we consider a second mode of star formation, \ie Population~III (Pop.~III). The transition between these two modes occurs at a metallicity of $\SI{e-6}{\Zsun}$ where it is expected that high-density gas clouds can fragment via dust processes \citep{Omukai2005}. Flagging of star formation sites and the calculation of the star formation efficiency are similar for both star formation modes, but below this metallicity threshold, individual Pop.~III stars (rather than an integrated stellar population) are drawn from an IMF following:
\begin{equation}
    \frac{\mathrm{d} N}{\mathrm{d}\log M}\propto M^{-1.3}\exp\left[-\left(\frac{M_{\rm char}}{M}\right)^{\!\!1.6}\right] \, ,
    \label{eq:p3imf}
\end{equation}
where $M_{\rm char} = \SI{100}{\Msun}$ as in \cite{Wise2012,Kimm2017,Katz2023-popiii}. Only a single Pop.~III star is allowed to form at each time step. 

\subsubsection{Jeans Inspired Star Formation}
One of the key characteristics of the turbulence-inspired star formation criteria is that gas can reach very high densities $>\SI{e5}{cm^{-3}}$ as star formation is often held up by the inferred turbulent pressure. This particularly manifests in cosmological simulations where emission line tracers (\eg [C~{\small III}]~$\lambda\lambda$1908) probe gas up to $\SI{e5}{cm^{-3}}$ \citep[e.g.][]{spdrv1} and it is not clear whether such high gas densities are a general characteristic of high-redshift galaxies (\citealt{Isobe2023}, although c.f. \citealt{Senchyna2024,Topping2024,Topping2024c}). Moreover, the terminal momentum from SN explosions scales weakly with gas density \citep[e.g.][]{Thornton1998} so that if the gas is artificially too high in the simulation, SN are less effective. For this reason, we consider a model with boosted star formation efficiency in high-density regions: when the Jeans length is unresolved on the highest level of refinement, we artificially increase the efficiency ($\epsilon_*$) defined in Equation~\ref{eq:sfe} by a factor
\begin{equation}
    f=\frac{n_{\rm H}}{\SI{100}{cm^{-3}}}+1,
    \label{eq:jeans_boost}
\end{equation}
where $n_{\rm H}$ is the gas number density in $\si{cm^{-3}}$. The value of 100~cm$^{-3}$ is chosen as the lower limit of the enhancement as this represents the typical densities of GMCs in the local Universe \citep[e.g.][]{Chevance2020}. It also represents the typical density where the Jeans length (not turbulent Jeans length) is unresolved at the finest refinement level\footnote{For neutral hydrogen at \SI{100}{K}, at a density of \SI{100}{cm^{-3}}, the Jeans length is $\sim \SI{10}{pc}$ which is approximately the physical resolution of our simulation}. Recent simulations that adopt a sink particle algorithm for star formation show that the efficiency can be higher than what is inferred from the multi-freefall method \citep{Kang2024}.

\subsubsection{Strömgren Turbulence Inspired Star Formation}
One of the key issues with radiation transfer simulations is that for typical star formation efficiencies $\lesssim\SI{10}{\percent}$, the Strömgren sphere is unresolved (see discussion in \eg \citealt{Rosdahl2015b}). This means that photoionization, radiation pressure, and photoheating are all unresolved which can have significant numerical impacts \citep[e.g.][]{Deng2024a}. Furthermore, corrections are often made in post-processing to measure emission line luminosities when such a scenario arises \citep[e.g.][]{Smith2022, spdrv1}. 

To address this, we can require that our star formation algorithm spawns stellar particles as to always locally resolve the Strömgren sphere. The Strömgren radius (\citealt{Stromgren1939}) is computed as 
\begin{equation}
  \label{eq:stromgrensphere}
  r_{\text{S}} = \left(\frac{3\, Q}{4\, \pi n_{\rm H}^{2} \alpha_{B}}\right)^{1/3} \, .
\end{equation}
Here $Q$ is the production rate of hydrogen-ionizing photons (\ie Lyman-continuum) that directly scales with the stellar particle mass, and $\alpha_{B}$ is the Case~B recombination rate of hydrogen at \SI{e4}{K}. 

\begin{figure}
    \centering
    \includegraphics[width=0.45\textwidth]{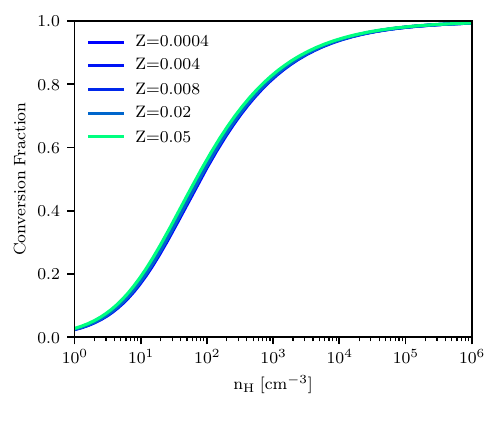}
    \caption{Fraction of the gas cell mass that needs to be converted into zero-age main-sequence star particles at different metallicities (different colored lines) such that the Strömgren sphere is resolved. For this figure, we assume a spatial resolution of 5.34~pc, which is the maximum level of refinement at $z=13.3$ for the fiducial simulations presented below. In the code this will change depending on redshift to correspond to the physical resolution of the cells.}
    \label{fig:strom_eff}
\end{figure}

In Figure~\ref{fig:strom_eff}, we show the fraction of the gas cell mass that needs to be converted into star particles so that $r_{\rm S}>\Delta x/2$ for SEDs of varying metallicity from {\small Starburst99} \citep{Leitherer1999}. We note that these curves are only weakly dependent on SED. At a density of $\SI{100}{cm^{-3}}$, more than \SI{50}{\percent} of the gas cell must be converted into star particles for the Strömgren sphere to be resolved. Such efficiencies are rarely reached in our simulations leading to a scenario where $r_{\rm s}$ can be catastrophically under-resolved.    

For this reason, we develop a third star formation algorithm that differs from the fiducial model in two ways. First, we include a more extreme boost on the star formation efficiency such that
\begin{equation}
    f=\left(\frac{n_{\rm H}}{\SI{100}{cm^{-3}}}+1\right)^2.
    \label{eq:strom_boost}
\end{equation}
In this case, the star formation efficiencies are often $\gg1$, leading to star formation faster than the free-fall time (see e.g.\ \citealt{Kang2024}). Second, we similarly draw from a Poisson distribution the number of star particles to form; however, if the number of particles is fewer than that needed to resolve the Strömgren radius, we do not allow the cell to form a particle. While the motivation for this algorithm is purely numerical, the effect is such that the physics of H~{\small II} regions is better resolved than in our fiducial model, at the expense of star particles becoming exceptionally massive (\ie star particles have masses $\gtrsim \SI{e5}{\Msun}$ when formed in dense gas at $n_{\rm H}>\SI{e4}{cm^{-3}}$ when the resolution is \SI{10}{pc}). 

\subsubsection{Density Threshold Star Formation}
Although the turbulent star formation recipes are calibrated on much higher-resolution simulations where turbulence dictates the efficiency of star formation, in our case, turbulence is measured on the grid scale. Thus, our computed turbulent properties might not reflect the more physical conditions when smaller scales are resolved or more sophisticated schemes that track turbulence on subgrid scales \citep[e.g.][]{Semenov2024}. For this reason, we consider a more traditional criterion for star formation such that gas with $n_{\rm H}>\rho_{\rm sf}$, where $\rho_{\rm sf}$ is the density threshold for star formation, and $T<\SI{500}{K}$ are eligible for star formation with no other conditions imposed. We consider two efficiencies of \SI{10}{\percent} \citep[e.g.][]{Grisdale2018} and \SI{100}{\percent} \citep[e.g.][]{Hopkins2014} and two density thresholds of $\SI{e2}{cm^{-3}}$ and $\SI{e3}{cm^{-3}}$.

\subsection{Radiation Transfer}
Multi-frequency radiation transfer (RT) is computed on-the-fly in the simulation in eight energy bins spanning the IR to the EUV, following the methods presented in \cite{Rosdahl2013}. Each of the eight photon groups and its purpose are listed in Table~\ref{tab:engy_bins}. Photons are advected using the M1 moment method \citep{Levermore1984} and we employ a GLF scheme to construct intercell fluxes. To reduce the computational demands of the RT, we employ a reduced speed of light approximation setting $c_{\rm sim}=0.01 \, c$ and subcycle \citep{Commercon2014} the radiation (and cooling) steps up to 500 times per each hydrodynamic time step. In practice, once SN explosions occur, the number of sub-cycles is typically $1-2$. In addition to photoionization and photoheating, we consider the dynamical effects of radiation pressure in the single scattering limit for UV and optical photons and multiple scatterings in the IR \citep{Rosdahl2015}. 

For a subset of models, we also consider the impact of Ly$\alpha$ radiation pressure following \cite{Kimm2018}\footnote{Note that this model underestimates the force multiplier by a factor of $\sim 3$ compared to analytical estimates \citep[e.g.][]{Lao2020,Nebrin2025}, but overestimates the impact of Ly$\alpha$ pressure by a factor of $\sim 7$ due to employing a point source approximation. In this respect, the approach of \cite{Kimm2018} may be considered as an upper bound (see further discussion below).}. Briefly, we calculate the number of recombinations in each cell that are expected based on density, temperature, and ionization fraction as well as the fraction of the recombinations that produce Ly$\alpha$ photons. We then estimate the multiplication factor ($M_{\rm F}$) based on the local gas column density and metallicity. The net force on the gas is then $F_{\rm Ly\alpha}=M_{\rm F}L_{\rm Ly\alpha}/c$, where $L_{\rm Ly\alpha}$ is the Ly$\alpha$ luminosity and $c$ is the speed of light. 

\begin{table}
    \centering
        \caption{Eight photon groups used for radiative transfer in this work. Listed are the photon group names, their lower and upper energies, and their main physical effects. }
    \begin{tabularx}{\columnwidth}{lllX}
    \toprule
    Group Name & $E_{\rm low}$ & $E_{\rm high}$ & Function \\
    & (eV) & (eV) & \\
    \midrule
    IR & 0.1 & 1.0 & Infrared radiation pressure \\
    Optical & 1.0 & 5.6 & Direct radiation pressure \\
    
    FUV & 5.6 & 11.2 & Photoelectric heating, Mg~{I}, Si~{I}, S~{I}, Fe~{I} ionization \\
    
    LW & 11.2 & 13.6 & H$_2$ \& CO dissociation, C~{I} ionization \\
    
    EUV$_1$ & 13.6 & 15.2 & H~{I}, N~{I}, O~{I}, Mg~{II} ionization \\
    
    EUV$_2$ & 15.2 & 24.59 & H$_2$, C~{II}, Si~{II}, S~{II}, Fe~{II}, Ne~{I} ionization \\
    
    EUV$_3$ & 24.59 & 54.42 & He~{I}, O~{II}, C~{III}, N~{II}, N~{III},  Si~{III}, Si~{IV}, S~{III}, S~{IV}, Ne~{II}, Fe~{III} ionization \\
    
    EUV$_4$ & 54.42 & $\infty$ & He~{II}, O~{III}+, N~{IV}+, C~{IV}+, Mg~{III}+, S~{V}+, Si~{V}+, Fe~{IV}+, Ne~{III}+ ionization \\
    \bottomrule
    \end{tabularx}
    \label{tab:engy_bins}
\end{table}

There are two sources of radiation in the simulation: star particles and a UV background. Pop.~III stars inject radiation based on their mass following \cite{Schaerer2002}, while the amount of radiation from Pop.~II stars is calculated by interpolating tables computed with {\small STARBURST99} \citep{Leitherer1999} as a function of age, metallicity and initial mass function. These luminosities are scaled based on the stellar particle mass. In our fiducial model, we adopt an IMF with an upper-mass cutoff of $\SI{120}{\Msun}$, an upper-mass slope of $-2.3$, a break mass of $\SI{0.5}{\Msun}$, a lower cutoff mass of $\SI{0.08}{\Msun}$, and a low mass slope of $-1.3$. Note that we do not include a sub-grid escape fraction, \ie we inject the exact amount of radiation as prescribed by {\small STARBURST99}. In future work, we will consider variations on this fiducial IMF.

To account for the radiation of galaxies and quasars from regions outside the Lagrange (zoom-in) volume of our selected halo, we also include a (non-advected) UV background \citep{Haardt2012}, considering only LyC ($>13.6$ eV) radiation. We exponentially suppress this background field at $z>5.3$ to model a late cosmic reionization \citep[e.g.][]{Kulkarni2019}. In this work, all simulations are stopped at $z=6$ and the UV background thus plays a limited role, but we will pursue simulations to lower redshifts in future work. Furthermore we exponentially suppress the background at $n_{\rm H} > \SI{100}{cm^{-3}}$ to account for the self-shielding of gas in dense environments \citep{Rosdahl2012}. Note that we do not follow the ionizing radiation that can be emitted by free-bound recombination and thus assume that it is all absorbed locally (the on-the-spot approximation).

To calculate the photoionization and photoheating rates, the mean energy-weighted and number-weighted cross sections must be known for each species. This depends on the SED of each star particle, but is numerically costly to perform this calculation locally for each individual cell. Instead, we compute the mean SED to estimate the cross sections. At very high redshifts where star formation is dominated locally by individual Pop.~III events, we recompute cross sections at every coarse time step as a luminosity-weighted mean over all star particles since stellar evolution proceeds rapidly for massive stars. At lower-redshift, we perform this calculation every ten coarse time steps\footnote{In practice, we wait for the Pop.~III star formation to become subdominant compared to Pop.~II and update this parameter when restarting the simulation from a checkpoint.}. Finally, to ensure that the radiation phase is well sampled in the simulation we ensure that once star particles form, fine time steps are a maximum of \SI{e5}{yr}. In practice, once SNe explode, the fine time steps on the maximum level of refinement are of $\mathcal{O}(\SI{100}{yr})$.

\subsection{Stellar Feedback \& Chemical Enrichment}

\subsubsection{Pop.~III}
We assume that Pop.~III stars can end their lives via four channels based on the mass of the star particle: core-collapse SN, hypernova, direct collapse to a black hole, or as a Pair Instability SN (PISN). In cases where Pop.~III stars explode, yields are adopted from \cite{Umeda2002,Nomoto2006,Nomoto2013}. Pop.~III stars between $10-\SI{20}{\Msun}$ explode as core-collapse SN injecting the equivalent of \SI{e51}{erg}. In the mass range $20-\SI{40}{\Msun}$, Pop.~III stars explode as hypernova with injection energies of $10-\SI{40e51}{erg}$ based on mass following \cite{Nomoto2006}. We assume that Pop.~III stars in the mass range $40-\SI{140}{\Msun}$ directly collapse to black holes injecting no metals or energy. Finally, Pop.~III stars in the mass range $140-\SI{300}{\Msun}$ explode as PISN with energy based on the remnant He core following \cite{Heger2002}. Whenever a SN explodes, we ensure that the expected final terminal momentum of the explosion is captured, independent of resolution (see below). Pop.~III main sequence lifetimes are adopted from \citep{Schaerer2002}.

\subsubsection{Pop.~II}
We base the injection of momentum, energy, and metals from Pop.~II stellar populations on the \textsc{vintergatan} model (described in detail in \citealt{Agertz2021} and \citealt{Rey2023Vintergatan}; see also \citealt{Agertz2013, Agertz2015, Agertz2020}). The model includes core-collapse SN, type Ia SN, and stellar winds. This model leads to realistic star formation modes across cosmic times (\citealt{Renaud2022, Segovia2022}) and is validated against $z=0$ observables, producing stellar masses compatible with abundance matching expectations for Milky-Way mass dark matter halos (\eg \citealt{Rey2023Vintergatan}) and discs whose structure resembles our Milky-Way's thin and thick disk \citep{Agertz2021, Renaud2021VG2, Renaud2021VG3}. 

For each stellar population of a given age, we stochastically sample individual SNeII progenitors exiting the main sequence during each time step, following the main-sequence lifetime-mass relation from \cite{Schaerer1993}. We assume that stars with $8 \leq m_{\star}/{\si{\Msun}} \leq 25$ explode as SNeII with an energy of $E_{\rm SN} = \SI{e51}{erg}$, consistent with the adopted yields model (described below). For a subset of models, we allow Pop.~II stars with mass $\geq \SI{25}{\Msun}$ to explode as hypernova (HN) following the model of \cite{Nomoto2006} where the injected energy, metal yields, and fraction of stars that explode as HN evolve with stellar mass and metallicity.

\vspace{2cm}
\subsubsection{Energy and Momentum Injection from Supernova}
We always inject momentum from the initial ballistic ejecta scaled as $\sqrt{2 \, E_{\mathrm{SN}} / M_{\text{ej}}}$, where $M_{\text{ej}}$ is the mass loss associated with a progenitor of $m_{\star}$ according to \citet{Limongi2018}. When the supernova cooling radius, computed as 
\begin{multline}
    r_{\mathrm{cool}} = 30 \, \left(\frac{n}{1 \, \mathrm{cm}^{-3}}\right)^{-0.42} \times \\ \, \left({\rm Max}(Z_{\mathrm{gas}}, 0.01Z_{\odot})\right)^{-0.18} \, \left(\frac{E}{E_{\mathrm{SN}}}\right)^{0.2857} \ \si{pc},
\end{multline}
is resolved by at least four grid cells, we additionally inject \SI{e51}{erg} of thermal energy as the Sedov-Taylor phase should be resolved. If the cooling radius is unresolved, we inject the terminal momentum according to 
\begin{multline}
p = 2.95 \, \left(\frac{E}{E_{\rm SN}}\right)^{0.941} \, \left(\frac{n}{1 \, \mathrm{cm}^{-3}}\right)^{-0.1176}\times \\ \, ({\rm Max}(Z_{\mathrm{gas}}, 0.01Z_{\odot}))^{-0.2} \, {\rm M_{\odot}} \, \mathrm{km \, s}^{-1},
\end{multline}
(see \eg \citealt{Blondin1998} and \citealt{Kim2015}). All injections of mass, momentum, and metals (see below) are performed in the local 8 grid cells surrounding the stellar particle. For numerical reasons, we apply a limiter such that supernova ejecta cannot exceed \SI{4000}{km.s^{-1}} and the gas cannot be heated above \SI{e9}{K}. 

\begin{figure}
    \centering
    \includegraphics[width=0.45\textwidth]{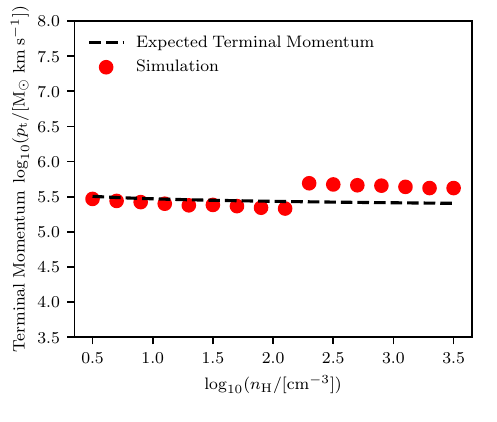}
    \caption{Expected terminal momentum from SN explosions (dashed black line) as a function of gas density compared to that measured in idealized MEGATRON simulations (red circles). The discontinuity at $n_{\rm H}=10^{2.3}\,{\rm cm^{-3}}$ represents the transition point between a resolved and unresolved cooling radius for the SN remnant. The expected terminal momentum follows the results of \protect\cite{Kim2015}. }
    \label{fig:term_mom}
\end{figure}

In Figure~\ref{fig:term_mom}, we show the terminal momentum recovered in an example simulation where we launch a SN in a homogeneous box of varying density. At the lowest densities, where the cooling radius is well resolved, we recover the expected value. As the density increases, we begin to under predict the momentum due to the worsening resolution. For this particular setup, the transition between energy and momentum injection occurs at $n_{\rm H}\sim \SI{125}{cm^{-3}}$ at which point we slightly overshoot the terminal momentum due to the sharp transition. We then begin to converge back to the expectation value as density increases. The mean absolute deviation from the expectation value is \SI{40}{\percent} and adopting a smoother transition \citep[e.g.][]{Kimm2014} would marginally improve this behavior (see also \citealt{Hopkins2018}).

\subsubsection{Metal Yields}
Each SNeII further injects individual elements according to the progenitor-mass and metallicity dependent grid of \cite{Limongi2018}. We modify the yields to account for the radioactive decay of specific isotopes on timescales much shorter than galactic evolution, namely (i) the decay half-life, $t_{1/2}$ of $^{22}$Na into $^{22}$Ne is \SI{2.6}{yr} and its contribution is thus added to Ne yield; (ii) the contribution of $^{26}$Al is added to Mg ($t_{1/2}= \SI{0.73}{Myr}$); and (iii) the contributions of $^{56}$Ni and $^{56}$Co are added to the Fe yield ($t_{1/2} \leq \SI{100}{days}$). Stars with $m_{\star} \geq \SI{25}{\Msun}$ are assumed to directly collapse without exploding, but still inject metals into the ISM following their mass-loss due to winds \citep{Limongi2018}. 

The yields from \cite{Limongi2018} already include the metal production from pre-supernova O and B stellar winds. Compared with \citealt{Agertz2013, Agertz2015}, we do not account for the momentum injection of these winds which are expected to be a subdominant feedback channel given the low-metallicities considered here. We however account for the mass, momentum and metal injection from asymptotic giant branch (AGB) stars, computing their IMF-averaged mass and momentum contributions as in \citet{Agertz2013} and using yields from \citet{Ritter2018}.

We also account for SNeIa at a rate of $2.6\times10^{-13}\ {\rm SNIa\ yr^{-1}\ M_{\odot}^{-1}}$, assuming the delay-time distribution of \citet{Maoz2017}, injecting metals (and thermal energy or momentum depending on resolution) according to \citet{Seitenzahl2013}.

\section{Regulating Star Formation}

To explore how the MEGATRON model regulates high-redshift galaxy formation, we perform cosmological zoomed simulations of a high-redshift dwarf galaxy ($M_{\star} \approx \SI{e7}{\Msun}$ at $z=6$). In this section, we present results from our fiducial implementation (as defined in Table~\ref{tab:sim_parameters}), before performing a wide exploration of numerical parameters and physics modules in Section~\ref{sec:results:numerics}.

The initial conditions for our target dwarf galaxy represent a halo that grows to a virial mass of $\sim \SI{e9}{\Msun}$ at $z=6$. The halo was selected from a parent volume of $5^3\,\si{Mpc^3}$ that was run with $128^3$ dark matter particles. Higher resolution initial conditions were generated with two additional levels of refinement within the Lagrangian region of the halo such that the halo remains uncontaminated by low-resolution dark matter particles well beyond its virial radius by $z=6$. The highest resolution dark matter particles have a mass of \SI{2.1e4}{\Msun.\hred^{-1}} and we assume a flat cosmology consistent with \cite{Planck2020} such that $H_0=\SI{67.66}{km.s^{-1}.Mpc^{-1}}$, $\Omega_{\rm m}=0.3111$, $\Omega_{\rm b}=0.04897$. We allow for seven additional levels of refinement beyond what is present in the initial conditions such that we reach a spatial resolution of $\sim11$ physical pc at $z=6$.

\begin{figure}
    \centering
    \includegraphics[width=0.45\textwidth]{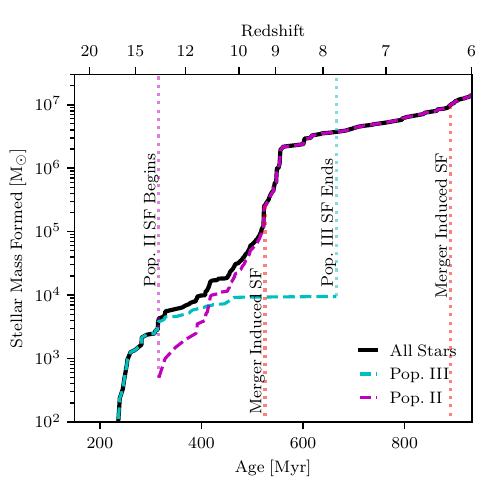}
    \includegraphics[width=0.45\textwidth]{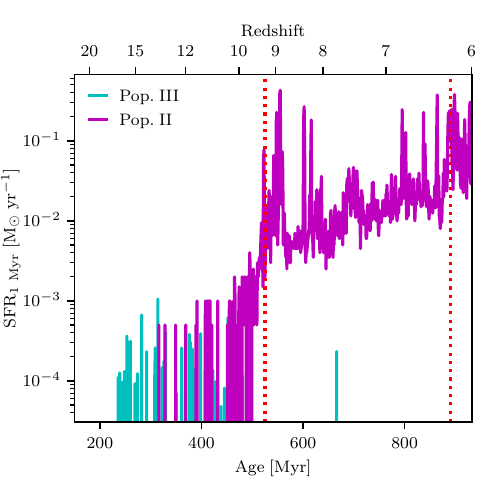}
    \caption{Cumulative stellar mass formed (top) or \SI{1}{Myr}-averaged star formation rate (bottom) as a function of time for all stars in the simulation. Pop.~III star formation (cyan) begins first and continues until $z\approx8$, but is rapidly overtaken by Pop.~II star formation (purple). Rapid increases in stellar mass (vertical red dashed) track vigorous assembly through major mergers.}
    \label{fig:fid_SFH}
\end{figure}

In Figure~\ref{fig:fid_SFH}, we plot the cumulative stellar mass growth within the full Lagrangian region as a function of the age of the universe and redshift down to $z=6$. Metal-free star formation begins at $z=16.4$ with the formation of a \SI{111}{\Msun} Pop.~III star. Due to its mass, this star collapses directly into a black hole. Subsequent Pop.~III stars form and by $z=15.4$, the first Pop.~III SN explodes, polluting the gas with the first heavy elements. By $z=13.4$, the halo recovers from the SN explosion and the first metal-enriched Pop.~II stellar cluster forms. Pop.~III star formation continues until $z\sim7.7$ but remains well sub-dominant compared to Pop.~II star formation at $z<10$.

\begin{figure}
    \centering
    \includegraphics[width=0.45\textwidth]{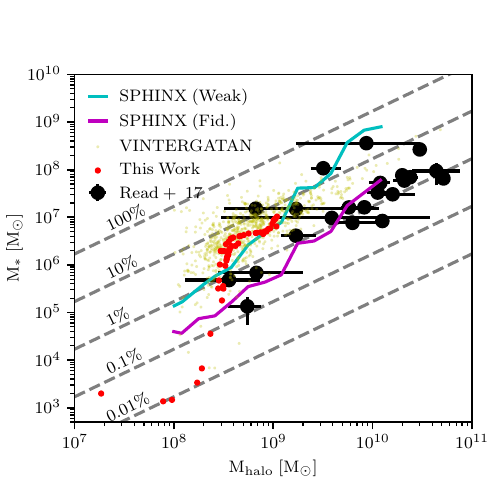}
    \caption{Stellar mass-halo mass relation for the main halo between $z\sim15-6$ (red points). Dashed black lines show contours of constant efficiency defined as ${\rm M_*/\frac{\Omega_b}{\Omega_{DM}}M_{halo}}$. Throughout the simulation, our main halo aligns with the weak feedback model from the {\small SPHINX} simulation (cyan) rather than their fiducial relation (purple) which was necessary to reproduce the $z=6$ UV luminosity function. Nonetheless, the simulation is higher but broadly aligned with $z=0$ dwarf galaxies (black points; \protect\citealt{Read2017}) and $z=6$ data from the {\small VINTERGATAN-GM} simulations \citep{Rey2023Vintergatan} on which our feedback model is based. This highlights that successful low-redshift galaxy formation models (e.g.\ VINTERGATAN) do not guarantee well-regulated high-redshift galaxies, warranting an in-depth exploration of numerical parameters in this regime (Figure~\ref{fig:reg_all}).}
    \label{fig:fid_smhm}
\end{figure}

The star formation history of the halo is very ``bursty''. Large fluctuations of nearly two orders of magnitude in SFR occur on Myr timescales (see bottom panel of Figure~\ref{fig:fid_SFH}). The first burst occurs with the formation of the first Pop.~III stars and the SFR slows down after the first Pop.~III SN event. Another strong burst of star formation occurs at $z\sim9.3$ which is induced by a major merger. Similar behavior is seen at $z\sim6.2$ where two smaller haloes initiate a three-way merger, which again induces star formation.

To understand how well the feedback is regulating star formation, we use the {\small ROCKSTAR} halo finder \citep{Behroozi2013} to identify haloes in the simulation. We then assign star particles to haloes by selecting all star particles within the halo virial radius. For stars particles that are within the virial radius of more than one halo (which can happen during mergers or when there are subhaloes), we assign them to the halo that minimizes $d/r_{\rm vir}$, where $d$ is the distance to the center of each halo.

In Figure~\ref{fig:fid_smhm}, we plot the stellar~mass-halo~mass relation for the main halo for all snapshots after first star formation. We compare them to the inferred properties of $z=0$ dwarf galaxies from \cite{Read2017}, simulated galaxies at $z=6$ from the {\small SPHINX} simulation that was designed to study the high-redshift Universe, and simulated galaxies at $z=6.1$ in the Lagrangian patches of the \textsc{vintergatan} project that agrees well with data from the low-redshift Universe (\citealt{Rey2023Vintergatan}, Rey et al. in prep.).   

With the fiducial model, our dwarf galaxy at $z=6$ broadly consistent with but perhaps slightly higher than the average stellar mass at a given halo mass of local dwarf galaxies at $z=0$. This is to be expected due to the mismatch in timescales and physical mechanisms that govern star formation in the two redshift regimes -- dwarf galaxies with stellar masses of $\approx \SI{e6}{\Msun}$ at $z=0$ are likely to have been strongly affected by reionization quenching over their history \citep[e.g.][and references therein]{Efstathiou1992, BenitezLlambay2020,Rey2020, Katz2020}. This would lead to a sharp decline (or altogether flattening) in their stellar mass growth rate for a given halo mass unit, compared to our high-redshift dwarf for which reionization quenching has yet to take effect. 

Nevertheless, the elevated stellar mass is worrying, particularly since it follows the `Weak feedback' model from the {\small SPHINX} simulations. The `Weak feedback' model from the {\small SPHINX} simulation over-predicts the UV luminosity function of high-redshift galaxies \citep{Rosdahl2018}. In contrast, our test dwarf galaxy is compatible with the \textsc{vintergatan} data at $z=6.1$. The numerics of the feedback implementation for SNII, SNIa, and winds are nearly identical between \textsc{vintergatan} and this work, but we stress that this agreement is non-trivial due to the vastly different heating and cooling physics (equilibrium vs.\ non-equilibrium + radiative transfer) and numerical resolutions.

From these comparisons, we conclude that:
\begin{enumerate}
 \item Our fiducial model may require additional feedback energy and/or channels to regulate high-redshift star formation, as is successfully done in \eg the {\small SPHINX} `Fiducial' model. Alternatively, a model where a higher fraction of the stellar particle mass evolves off the main sequence so that the mass-to-light ratio decreases at later times (e.g.\ a top-heavy IMF) would achieve a similar `apparent' regulation. 
 \item Successful low-redshift models such as \textsc{vintergatan} do not guarantee a well-regulated star formation at $z>6$, and that such lack of regulation at high redshift does not imply over-massive galaxies at low redshift. 
\end{enumerate}
As stellar masses are difficult to infer at high-redshift \citep[e.g.][]{Narayanan2024}, it remains unclear what the true stellar mass-halo mass relation is at high-redshift.

Direct constraints on the ISM conditions of high-redshift galaxies from JWST are one potential route to reconcile these high- and low-redshift successes together. The remainder of this paper thus focuses on how variations in subgrid modeling affects the regulation of star formation (Section~\ref{sec:results:numerics}), combined with an in-depth discussion of their associated spectroscopic observables at $z>6$ (Section~\ref{sec:discussion:jwst}).

\begin{table*}
\centering
        \caption{List of parameters used for our 25 simulations. M$_*$ is the stellar mass resolution, $\varepsilon_\mathrm{ff}$ is the star formation efficiency per free-fall time, $t_\mathrm{SN}$ is the time to the first supernova, $E_\mathrm{SN}$ is the energy per SNeII, HN refers to hypernovae being turned on or not, Ly$\alpha$ refers to Lyman-$\alpha$ radiation pressure being turned on or not, $M_\mathrm{DM}$ is the DM mass resolution, $l_\mathrm{max}$ is the maximum level of refinement, $c_\mathrm{sim.}$ is the value of the speed of light in the simulation. See the text for further details.}
    \begin{tabular}{
        p{3.4cm}
        >{\raggedleft\arraybackslash}p{.9cm}
        p{1.35cm}
        >{\centering\arraybackslash}p{.7cm}
        >{\centering\arraybackslash}p{1cm}
        p{.5cm}
        p{.5cm}
        p{.6cm}
        >{\centering\arraybackslash}p{1cm}
        >{\centering\arraybackslash}p{1cm}
        >{\centering\arraybackslash}p{.8cm}
        >{\centering\arraybackslash}p{.8cm}
        p{1cm}
    }
    \toprule
    Name &
    M$_*$ &
    $\epsilon_{\rm ff}$ &
    $t_{\rm SN}$ &
    $E_{\rm SN}$ &
    HN &
    Ly$\alpha$ RP &
    SF model &
    $\rho_{\rm SFR,min}$ &
    ${M_\mathrm{DM}}$ &
    $l_{\rm max}$ &
    ${c_{\rm sim.}}$ &
    Cooling model \\
         & (\si{\Msun}) & & (\si{Myr}) & (\SI{e51}{erg})  & &&& (\si{cm^ {-3}}) & (\SI{e4}{\Msun}) && ($c$) & \\
    \midrule
    Fiducial     & 500 & \Eq{eq:sfe} & 7 & $1$ & & & turb. & $10$ & $2.1$ &15 & 0.01 & PRISM \\
    $M_\star\times 2$   & 1000 & \Eq{eq:sfe} & 7 & $1$ & & & turb. & $10$ & $2.1$ &15 & 0.01 & PRISM \\
    $M_\star\times 8$   & 4000 & \Eq{eq:sfe} & 7 & $1$ & & & turb. & $10$ & $2.1$ &15 & 0.01 & PRISM \\
    $M_\star\times 20$   & 10000 & \Eq{eq:sfe} & 7 & $1$ & & & turb. & $10$ & $2.1$ &15 & 0.01 & PRISM \\
    Efficient & 500 & $\SI{100}{\percent}$ & 7 & $1$ & & & turb. & $10$ & $2.1$ &15 & 0.01 & PRISM \\
    Efficient, massive & 1000 & $\SI{100}{\percent}$& 7 & $1$ & & &turb.& $10$ & $2.1$ &15 & 0.01 & PRISM \\
    Efficient, more massive & 4000 & $\SI{100}{\percent}$ & 7 & $1$ & & & turb. & $10$ & $2.1$ &15 & 0.01 & PRISM \\
    $t_\mathrm{SN,0.5}$ & 500 & \Eq{eq:sfe} & 0.5 & $1$ & & & turb. & $10$ & $2.1$ &15 & 0.01 & PRISM \\
    $t_\mathrm{SN,1.0}$ & 500 & \Eq{eq:sfe} & 1.0 & $1$ & & & turb. & $10$ & $2.1$ &15 & 0.01 & PRISM \\
    $t_\mathrm{SN,2.0}$ & 500 & \Eq{eq:sfe} & 2.0 & $1$ & & & turb. & $10$ & $2.1$ &15 & 0.01 & PRISM \\
    $E_\mathrm{SN}\times 2$   & 500 & \Eq{eq:sfe} & 7 & $2$ & & & turb. & $10$ & $2.1$ &15 & 0.01 & PRISM \\
    $E_\mathrm{SN}\times 4$   & 500 & \Eq{eq:sfe} & 7 & $4$ & & & turb. & $10$ & $2.1$ &15 & 0.01 & PRISM \\
    $E_\mathrm{SN}\times 5$   & 500 & \Eq{eq:sfe} & 7 & $5$ & & & turb. & $10$ & $2.1$ &15 & 0.01 & PRISM \\
    HN     & 500 & \Eq{eq:sfe} & 7 & $1-40$ & $\checkmark$ & & turb. & $10$ & $2.1$ &15 & 0.01 & PRISM \\
    Ly$\alpha$     & 500 & \Eq{eq:sfe} & 7 & $1$ & &  $\checkmark$ & turb. & $10$ & $2.1$ &15 & 0.01 & PRISM \\
    Density     & 500 & 10\% & 7 & $1$ & & & dens. & $10^2$ & $2.1$ &15 & 0.01 & PRISM \\
    Jeans     & 500 & Eqs.~(\ref{eq:sfe}, \ref{eq:jeans_boost}) & 7 & $1$ & & & Jeans & $10$ & $2.1$ &15 & 0.01 & PRISM \\
    Strömgren     & 500 & Eqs.~(\ref{eq:sfe}, \ref{eq:strom_boost}) & 7 & $1$ & & & Ström. & $10$ & $2.1$ &15 & 0.01 & PRISM \\
    Efficient     & 500 & $\SI{100}{\percent}$ & 7 & $1$ & & & dens. & $10^2$ & $2.1$ &15 & 0.01 & PRISM \\
    Efficient, denser     & 500 &  $\SI{100}{\percent}$ & 7 & $1$ & & & dens. & $10^3$ & $2.1$ &15 & 0.01 & PRISM \\
    Efficient, denser, energetic   & 500 &  $\SI{100}{\percent}$ & 7 & $5$ & & & dens. & $10^3$ & $2.1$ &15 & 0.01 & PRISM \\
    Low-resolution     & 500 & \Eq{eq:sfe} & 7 & $1$ & & & turb. & $10$ & $16$ &14 & 0.01 & PRISM \\
    High-resolution    & 500 & \Eq{eq:sfe} & 7 & $1$ & & & turb. & $10$ & $0.26$ &16 & 0.01 & PRISM \\
    Fast-light     & 500 & \Eq{eq:sfe} & 7 & $1$ & & & turb. & $10$ & $2.1$ &15 & 0.1 & PRISM \\
    Simple cooling     & 500 & \Eq{eq:sfe} & 7 & $1$ & & & turb. & $10$ & $2.1$ &15 & 0.01 & Simple \\
    \bottomrule
    \end{tabular}
    \label{tab:sim_parameters}
\end{table*}

\begin{figure*}
    \centering
    \includegraphics[width=0.33\textwidth]{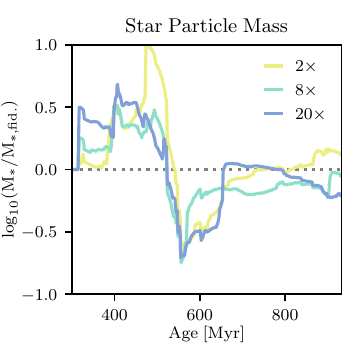}
    \includegraphics[width=0.33\textwidth]{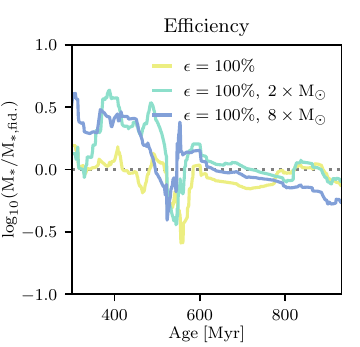}
    \includegraphics[width=0.33\textwidth]{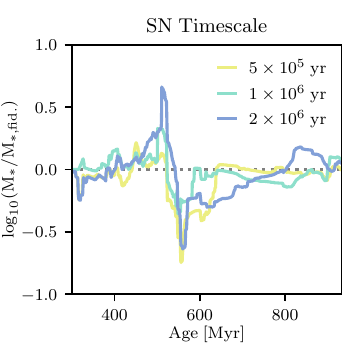}

    \includegraphics[width=0.33\textwidth]{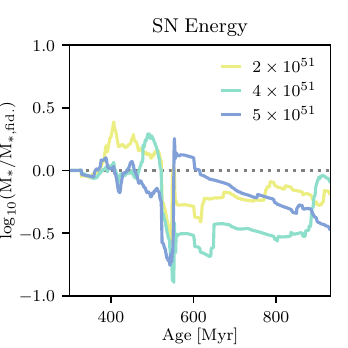}
    \includegraphics[width=0.33\textwidth]{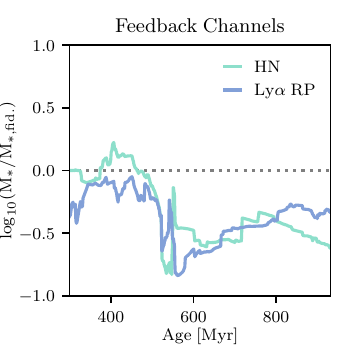}
    \includegraphics[width=0.33\textwidth]{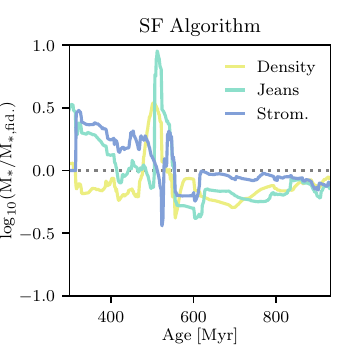}
    
    \includegraphics[width=0.33\textwidth]{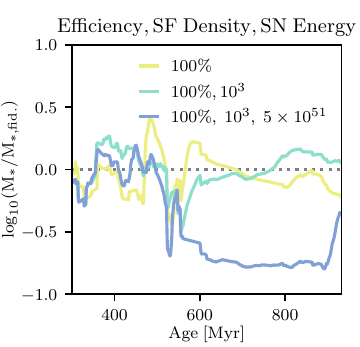}
    \includegraphics[width=0.33\textwidth]{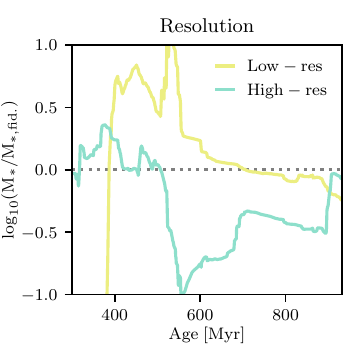}
    \includegraphics[width=0.33\textwidth]{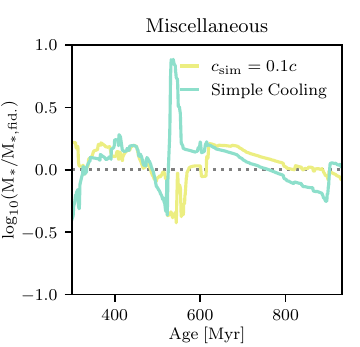}
    \caption{Ratio of cumulative stellar mass formed compared to our fiducial model for all numerical parameters explored in this paper. Panels show broad families of variations, with details of each model variation described in the text.}
    \label{fig:reg_all}
\end{figure*}

\section{Systematic variations of subgrid parameters} \label{sec:results:numerics}
In this Section, we vary several parameters of our fiducial model to understand how the stellar mass growth and star formation regulation responds to assumptions for our subgrid model. In total, we run 25 simulations and refer to Table~\ref{tab:sim_parameters} for a comprehensive list of the parameters varied in our simulation suite.

\subsection{Star Particle Mass}
Numerous simulations show that the clustering of SNe is key to drive strong galactic outflows, as coherent and repeated explosions create super-bubbles that break out from the disk and strongly enhance outflow loading factors \citep[e.g.][]{Keller2014,Yadav2017, Fielding2017, Gentry2019,Gentry2020,MCSmith2021,Keller2022_SN}. Increasing the stellar particle mass will lead to a larger cluster of SNe exploding at a similar location and time, rather than being spatially and temporally distributed in smaller-mass clusters. 

In the top left panel of Figure~\ref{fig:reg_all}, we show the ratio of stellar mass formed in simulations with star particle masses that are $2\times$, $8\times$, and $20\times$ that of the fiducial model. At $t=929$~Myr, corresponding to $z=6$, we see very little difference in the total stellar mass in any of the models. Increasing the star particle mass by factors of 2 or 8 provides no additional regulation. Increasing by a factor of 20 results in $\sim\SI{40}{\percent}$ lower stellar mass. Such a stellar mass is perhaps numerically ill-advised because it is $>\frac{\Omega_{\rm b}}{\Omega_{\rm dm}}m_{\rm dm}$, where $m_{\rm dm}$ is the dark matter particle mass. 

There does seem to be a systematic effect where prior to the major merger, the simulations with higher star particle mass overshoot the fiducial model in terms of stellar mass and then subsequently undershoot. The origin of this behaviour is unclear. The stellar mass of the fiducial simulation increases by $>100\times$ in the 150~Myr time-frame around the merger, so these variations are likely due to random fluctuations in star formation history.

\subsection{Star Formation Efficiency}
The typical star formation efficiencies per free-fall time measured by our fiducial star formation algorithm maximize at $\sim\SI{30}{\percent}$.
Increasing the efficiency can lead to star formation at lower densities and more clustered star formation. \cite{Hopkins2018} argues that this parameter has negligible effect on galaxy properties because star formation is feedback-regulated and gas will pile up until the same star formation is achieved, although the FIRE simulations adopt a fiducial value of \SI{100}{\percent}. In the top middle panel of Figure~\ref{fig:reg_all}, we show the ratio of stellar mass formed as a function of time for models where we ignore Equation~\ref{eq:sfe} and set the efficiency per free-fall time to \SI{100}{\percent}. Similar to what was found in \cite{Hopkins2018}, we find that the efficiency plays almost no role in changing the final stellar mass of the object at $z=6$. As discussed in the previous section, it is unclear what causes the runs with higher stellar mass to grow their stellar mass earlier compared to the fiducial model and this may be due to the stochasticity in the star formation history when the stellar mass is very low.

We further test whether combining an enhanced efficiency with a higher star particle mass ($2\times$ and $8\times$) can help regulate the galaxy. While the $2\times$ higher star particle mass results in no change in the final stellar mass, combining the high efficiency with an $8\times$ more massive star particle results in similar stellar mass suppression as increasing the star particle mass by a factor of 20. Nevertheless, this result is relatively modest given the extreme change in efficiency.

\begin{figure}
    \centering
    \includegraphics[width=0.43\textwidth,trim={0cm 0.2cm 0cm 0.4cm},clip]{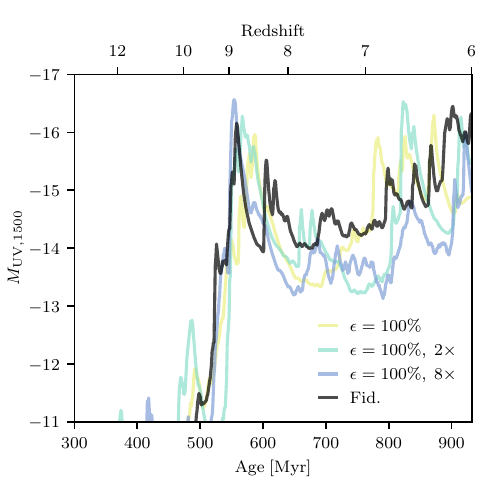}
    \includegraphics[width=0.43\textwidth,trim={0cm 0.2cm 0cm 0.4cm},clip]{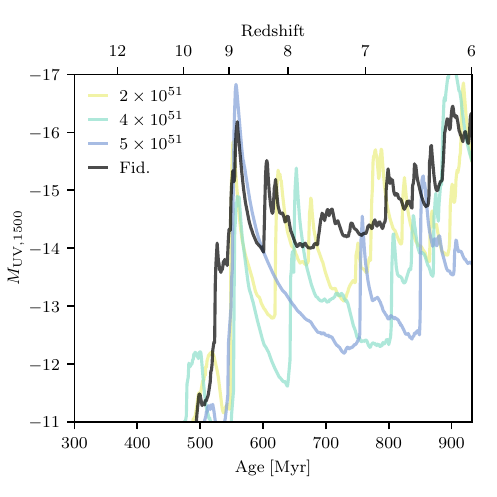}
    \includegraphics[width=0.43\textwidth,trim={0cm 0.2cm 0cm 0.4cm},clip]{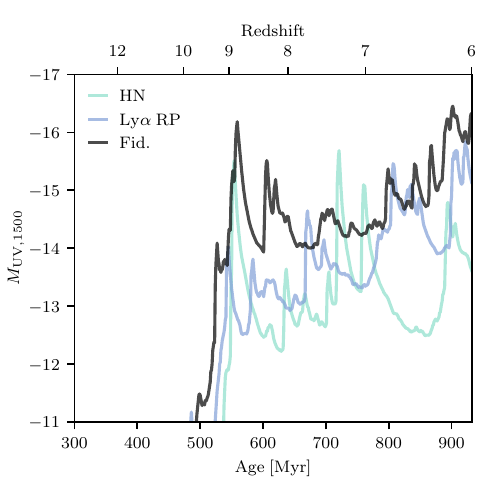}
    \caption{Intrinsic UV magnitude as a function of time for all stars in each simulation. In the top panel, we increase $\epsilon_{\rm ff}$ to \SI{100}{\percent} (yellow) or simultaneously increase $\epsilon_{\rm ff}$ and mass of the star particle by factors of 2 (green) or 8 (blue). Even with extremely efficient and clustered star formation, the halo gets at most \SI{0.4}{mag} brighter than our fiducial model (black). These parameters are thus unlikely to lead to large changes in the population-averaged UV luminosity function. The middle and bottom panels show the results for changing the strength of SN or adding in additional feedback channels.}
    \label{fig:MUV_eff}
\end{figure}

Although the integrated stellar mass does not change significantly, the higher efficiency should in principle lead to burstier star formation. This is interesting in the context of high redshift as UV variations driven by bursty star formation has been used as an explanation for the abundance of UV bright galaxies in the early Universe \citep[e.g.][]{Shen2023,Sun2023}. It should be noted that simulations do not agree upon how much $M_{\rm UV}$ fluctuates as a function of mass but a generic prediction is that as mass increases, UV magnitude fluctuations decrease \citep[e.g.][]{Kravtsov2024,Gelli2024,Shen2025}. In Figure~\ref{fig:MUV_eff}, we show the intrinsic 1500~\AA{} UV luminosity of the Lagrangian region as a function of time. The merger at $z\sim9$ leads to a star formation event which brings all simulations to a magnitude $\lesssim-16$. Even after a merger, where UV luminosity is expected to peak due to merger-induced star formation, increasing the star formation efficiency to \SI{100}{\percent} and increasing the stellar mass by a factor of 8 only leads to a modest increase of 0.4~magnitudes. This increase is well below the natural, intrinsic variability of M$_{\rm UV,1500}$. Hence changing the star formation efficiency in the context of our model is highly unlikely to lead to fundamental differences in the UV luminosity function. 

In summary, increasing the efficiency of star formation per freefall time and the stellar particle mass can have a minor impact on the final stellar mass of the object, although as we show below, these effects are well subdominant compared to other subgrid model parameters. 

\subsection{SN Timescale}
\label{sec:SN_time}
One of the key differences in our work compared to many other cosmological simulations is that only stars with mass $<\SI{25}{\Msun}$ explode as SN. While the typical \SI{100}{\Msun} evolves off the main sequence in $\sim\SI{3}{Myr}$, a \SI{25}{\Msun} star takes significantly longer ($\sim \SI{7}{Myr}$). This timescale has numerous implications for galaxy formation. \cite{Dekel2023} recently argued that when the free-fall time of the gas is significantly shorter than the feedback timescale\footnote{Note that in our case, this timescale would correspond to the SN timescale because stellar winds, radiation pressure and photoheating seem to have a limited impact at our resolution.}, star formation proceeds in a ``feedback-free'' mode, which can result in up to \SI{100}{\percent} of the gas converting into stars. If confirmed, such a model would provide a natural explanation for the abundance of bright galaxies observed at high-redshift without the need to invoke bursty star formation or a changing mass-to-light ratio. Due to the evolution of the ISM towards lower densities at lower redshift, this process is predicted to only operate in the early Universe. However, if the timescale to SN is much longer, the ``feedback-free'' process may continue to operate at progressively lower redshifts which could make the model inconsistent with observations. In the lower redshift Universe, the observed spatial de-correlation between H$\alpha$ and CO emission in disk galaxies is highly suggestive that star-forming clouds are rapidly dispersed \citep{Kruijssen2019}. Reproducing this behavior with simulations seems to require strong early feedback \citep{Semenov2021,Jeffreson2024} as matching the timescales with physically-based subgrid models is non-trivial \citep[e.g.][]{Keller2022,Hu2024}. Recent observations also support this picture \citep{Chevance2022}. Given the implications for feedback timescale in both the high- and low-redshift Universe, in this Section, we artificially modulate the time at which the first SN explode to understand its effect on early galaxy formation. 

For each star particle in the fiducial model, the first SN are eligible to explode at an age of $\sim\SI{7}{Myr}$. In practice, the age of the particle is often older than this value because we stochastically sample the IMF. To modify this behavior, we change the main-sequence lifetime-mass relation such that an \SI{8}{\Msun} star still explodes at the same time as in the fiducial model (34~Myr), but the \SI{25}{\Msun} main-sequence lifetime is reduced to either \SI{5e5}{yr}, \SI{1e6}{yr}, or \SI{2e6}{yr}, respectively. In this way we have elongated the period over which the SN explode, \ie potentially slightly reducing the spatial and temporal clustering of the SN, which seems to have little impact on our results, but the star-forming clouds are subject to feedback much earlier on in their evolution. 

The top right panel of Figure~\ref{fig:reg_all} shows the cumulative stellar mass formed of each of these three model variations, normalized to the fiducial model as a function of time. Despite the significant reduction in timescale to first SN, the integrated stellar masses do not change by $z=6$. Thus, if the total feedback budget remains unchanged, the time at which SN begin does not seem to impact how well the galaxy is regulated.  

We emphasize two key points. First, even if the star formation is not better regulated, this does not necessarily mean that the properties of the ISM are unaffected. For example, \cite{Keller2022} show how the properties of an isolated disk can change significantly using an empirical model for feedback injection where timescale is a key parameter. Second, even though the first SN explode earlier in these simulations, the time period over which the SN explode is effectively elongated. While stochastic sampling of the IMF is perhaps the most realistic, differences may be found if we explode all SN for the star particle at the exact same time \citep[e.g.][]{Bhagwat2024}. The results in this section do not conflict with the conclusions of these other studies.

\subsection{SN Energy}
\label{sec:SN_egy}
In order to better regulate early galaxy formation, \cite{Rosdahl2018} adopted the strategy of increasing the total number of SN by a factor of four. One could alternatively increase the total energy of a SN from the fiducial value of \SI{e51}{erg} to an arbitrarily higher value as is common technique in AMR simulations \citep[e.g.][]{Agertz2011,Semenov2021}. The final terminal momentum of a SN scales nearly linearly with SN energy and hence may lead to a significant reduction in star formation (compare the magenta and cyan lines in Figure~\ref{fig:fid_smhm} for results from {\small SPHINX}). 

Explanations for this modification can be both physical or numerical in origin. Stellar evolution models predict that not all SN explode with energies of \SI{e51}{erg} \citep[e.g.][]{Sukhbold2016}, with excursions to both lower and higher values. In the extreme scenario, some stars may explode as hypernovae (HN), which can have energies of $40\times$ the typically assumed value \citep{Nomoto2006}. This is further discussed in Section~\ref{sec:fb_chan}. Adopting a more top-heavy IMF can also lead to an increased number of SN, although the exact increase is highly dependent on the upper mass slope of the IMF. Making the IMF too top-heavy can also decrease the total number of SN for a stellar population as the mean mass of stars simultaneously increases as the high-mass slope flattens \citep{Prgomet2022}. This scenario is however intriguing for multiple reasons. First, modifying the IMF to be more top-heavy can both increase and decrease the SN energy budget which can impact the amount of regulation. Second the mass-to-light ratios are reduced at early times when massive stars are on the main-sequence and then increased at later times with respect to a standard Kroupa IMF, which could simultaneously help reconcile the abundance of early bright galaxies with the slow evolution in stellar mass growth towards lower redshifts (Katz~et~al.~{\it in~prep.}).

From a numerical perspective, although our SN scheme is designed to capture the final terminal momentum of a SN independent of resolution, energy can be artificially dissipated for other reasons. For example, the finite resolution means that shocks are much thicker in the simulation than in an idealized case. This has the effect of mixing the hot interior gas of the SN bubble with the denser gas outside the shell which can enhance cooling \citep[e.g.][]{Gentry2017,Gentry2019}. This is particularly bad in the context of non-equilibrium cooling as mixing ionized and neutral gas can lead to a strong enhancement in collisional excitation cooling compared to the case of collisional ionization equilibrium. This problem may be further exacerbated as the AMR grid tends to de-refine in SN-affected regions. The de-refinement is a diffusive process which again leads to mixing. For SN that drive strong outflows, the movement of the gas across the grid is also diffusive which could further lead to energy losses. Hence for the multitude of numerical reasons, energy may be artificially radiated even after the ``correct'' amount of momentum is injected into the ISM and represent variation of the classic ``over-cooling'' problem \citep{Katz1992}. Increasing the total energy of each SN explosion may help to combat these effects.

Finally, there remains some debate on the exact value of the final terminal momentum of a SN explosion and how it scales with SN energy, metallicity, and gas density. Certain models \citep[e.g.][]{Gentry2017} show increases of more than a factor of three in certain regimes compared to the fiducial model of \cite{Kim2015} used in this work. The model we adopt has been calibrated for individual SN; however, in the case where where multiple SN explosions occur near each other, there are also disagreements on whether the momentum is boosted with respect to the single SN limit for realistic environments \citep[e.g.][]{Gentry2019,Kim2017SB}.

With this discussion in mind, we run three simulations, testing how galaxy evolution changes when the energy per SN is increased to \SI{2e51}{erg}, \SI{4e51}{erg}, and \SI{5e51}{erg}, respectively. The cumulative stellar mass growth normalized by the fiducial model is shown in the left panel of the second row of Figure~\ref{fig:reg_all}. In all cases, the stellar mass at $z=6$ is lower than in the fiducial model with the $5\times$ boost model predictably producing the lowest integrated stellar mass of a factor of $\sim2$ lower than the fiducial model. However, the change in feedback energy seems to lead to non-trivial results. For example, just after the high-redshift merger, the model with the highest feedback energy temporarily exhibits the highest stellar mass (between $550-\SI{600}{Myr}$). This model suppresses star formation immediately preceding this event. We speculate that the gas has built up and the strong dynamical instabilities triggered by the merger lead to catastrophic star formation which cannot be regulated, even with $5\times$ boosted SN energy. A similar scenario seems to happen in the model with a $4\times$ boost in the SN energy. Comparing this model with the $2\times$ boost, they seem to track each others star formation history with the $4\times$ model having two times less stellar mass for most of the recent history of the galaxy. However, towards the second merger at $z\sim6.3$, it is likely that the excess gas reservoir leads to a stronger star formation event such that by $z=6$, the total stellar mass formed in the $4\times$ model is actually more than the $2\times$ model. We expect that if this simulation is run for longer, the model with higher SN energy will again fall below that of the fiducial and the $2\times$ boost model. Similar behavior was also observed in the simulation of \cite{Agertz2011} where the early suppression of star formation from stronger feedback was later compensated at lower redshift by a higher SFR due to the increased gas reservoir. More generally, the SFR decrease is expected to scale sub-linearly with boosts to feedback energy \citep[e.g.][]{Benincasa2016}.

\begin{figure}
    \centering
    \includegraphics[width=0.45\textwidth]{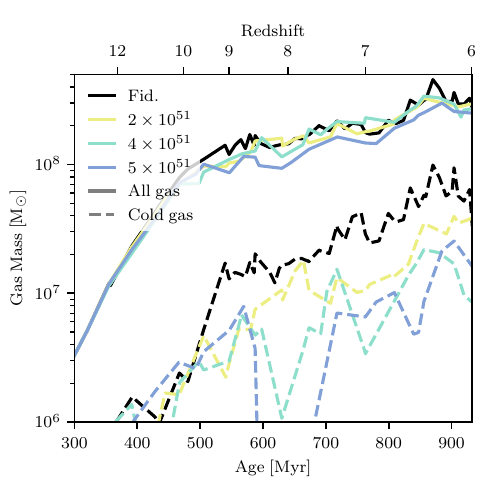}
    \caption{Total gas (solid lines) or cold ($T<\SI{500}{K}$) gas mass (dashed lines) within haloes as a function of redshift for the fiducial model (black) and models with varying SN energies (colors). Increasing feedback energy can diminish the available total and cold gas content in the galaxy, leading to a suppression of the final stellar mass (Figure~\ref{fig:reg_all}). }
    \label{fig:gas_fracs}
\end{figure}

While the gas reservoir argument is difficult to prove in practice, we can gain insight by looking at the evolution of the gas mass inside haloes as a function of redshift. In Figure~\ref{fig:gas_fracs}, we plot the total gas (solid lines) and the cold gas (dashed lines, defined as having a temperature $<\SI{500}{K}$) mass within haloes as a function of redshift. When the feedback energy is increased by a factor of $2-4$, the total gas reservoir within the haloes is almost the same as in the fiducial model while we find a slight decrease for a SN energy of $5\times$ the fiducial value. This is likely due to both ejective and preventative feedback and could explain why the stellar mass growth of the $5\times$ boost model is fundamentally different from the other simulations. However, between $z\sim10.5-9.5$, right at the time of the first merger, the model with the highest SN energy has temporarily more cold gas than the other boosted models. This provides an explanation as to why there is such a strong upturn in the cumulative stellar mass for the strongest feedback model at these redshifts in Figure~\ref{fig:reg_all}. The cold gas is then very strongly suppressed for a long duration leading to a decline in the average SFR. 

In all model variations, the amount of cold gas is decreased at $z<10$ compared to the fiducial model which helps explains the decrease in total stellar mass formed. However, because the gas is still within the haloes, continuous energy must be injected in order to prevent the gas from collapsing and forming stars. For a SN boost of 5$\times$, more gas is expelled entirely from the halo compared to the other models, as seen by an overall decrease in total gas mass in Figure~\ref{fig:gas_fracs}. For this specific model, up to $\sim50\%$ of the baryons can be ejected from the halo. Ejective feedback seems to be indicative of regulating star formation for this particular model over a prolonged periods of time; although it is difficult to differentiate the role of heating versus ejection or preventative feedback.

Mergers clearly complicate the relationship between SN energy and star formation regulation as do high gas inflow rates. Such a problem is rather unique to the high-redshift Universe where major gas-rich mergers and intense cold gas accretion are more common compared to the low-redshift Universe. For the time periods where the galaxy is evolving more secularly, the feedback energy more directly translates into a stellar mass suppression as seen in other simulations \citep[e.g.][]{Hopkins2018}. 

\subsection{Feedback Channels}
\label{sec:fb_chan}
While in Section~\ref{sec:SN_egy} we speculated why the energy per SN may need to be increased at high redshift for both physical and numerical reasons, here we consider two additional forms of feedback that operate on earlier timescales than our fiducial SN model. 

\paragraph{Hypernova Feedback} Stellar evolution models predict that some fraction of low-metallicity massive stars may explode with SN energies of $10-40\times$ that of a typical SN \citep{Nomoto2006,Nomoto2013}. The metal yields from these events are distinct from other evolutionary pathways and thus could provide a unique chemical signature that remains in low-redshift stellar populations and may even be necessary to explain low-redshift chemical abundance patterns in stars \citep{Kobayashi2006}. Recent reionization simulations of \cite{Bhagwat2024} demonstrated how the inclusion of HN\footnote{Note that the HN model of \cite{Bhagwat2024} is different from what is used here although the key principles of a large increase in SN energy and a strong metallicity dependence are present.} regulates star formation, particularly in the low-metallicity regime in the early Universe.  

Our model for HN feedback is very similar to \cite{Kobayashi2007}. For stars with mass $\geq \SI{20}{\Msun}$, we assume a mass-energy relation such that a mass of 20, 25, 30, and $>\SI{40}{\Msun}$ explode with energies of 10, 10, 20, and 40 times that of a typical SN\footnote{We interpolate energies between these masses.}. We also assume that the fraction of stars that explode as HN is strongly dependent on metallicity such that at a metallicity $\leq\SI{5}{\percent \Zsun}$, the HN fraction is \SI{50}{\percent} but at solar metallicity, the HN fraction is \SI{1}{\percent}.

In the middle panel of the second row of Figure~\ref{fig:reg_all}, we show the cumulative star formation history of the simulation with HN compared to the fiducial model. By $z=6$, we find significant regulation such that the total stellar mass is reduced by more than a factor of four. Purely based on energetics, this result is surprising. The mean SN energy in the run with HN is \SI{4.75e51}{erg}. Despite the lower average energy, the HN model regulates better than any of the models with the uniformly enhanced SN energies, and \SI{30}{\percent} better than the model with $5\times$ the energy per SN. This result is likely caused by two effects. First, there is clearly a difference between injecting a large amount of energy in a single SN event, as is done in the HN model, and spreading the energy out over multiple events, i.e. impulsivity of the feedback seems to matter. Most similar to this are simulations where realistic time delays are compared with injecting all the SN energy at one particular time \citep[e.g.][]{Kimm2015,Keller2022_SN,Bhagwat2024}. Depending on the exact setup, stellar masses, ISM conditions, and outflow rates can also be affected by a single versus distributed injection of energy. Second, because the HN feedback comes from massive stars, it happens early on in the life-cycle of the star-forming cloud and thus may disrupt the cloud earlier. Considering the results of Section~\ref{sec:SN_time} where we found little difference in star formation regulation with SN timescale, this latter point may be moot; however, the energy budget may simply be too low with the fiducial SN model to disrupt the dense, high-redshift clouds that form in the fiducial simulation. 

The primary issue with the HN model in the context of stellar mass regulation is that it is highly biased towards low-metallicity environments, by construction. While this reflects the fact that HN are not commonly observed in the low-redshift Universe and the chemical abundance patterns of local stars indicate that this process should not occur often at high metallicities \citep{Kobayashi2006}, the halo that we have chosen to model is relatively low mass. Due to the mass-metallicity relation, higher mass haloes tend to be more metal-enriched \citep[e.g.][]{Tremonti2004} and this relation is now being extended to the very early Universe \citep[e.g.][]{Curti2024}, albeit with significant scatter. Using our model, \cite{Choustikov2025} demonstrated that HN have a clear impact on the mass metallicity relation, particularly at the low-mass end. Our halo is not generally representative of the results one might find when applying the HN model to a wider sample of galaxy masses. For example, in full box cosmological simulations, \cite{Bhagwat2024} found that HN tend to suppress star formation early on, but as metal enrichment proceeded, significantly less regulation occurred. 

This result is however very interesting in the context of our model. Our feedback model is based on that used in \cite{Agertz2013,Agertz2021}, where they found that in simulations of Milky Way mass galaxies run to $z=0$, the galaxies were perfectly consistent with the low-redshift stellar mass-halo mass relation. Since the majority of metals and stars form after $z=6$ for the Milky Way mass haloes in {\small VINTERGATAN}, this early evolution is unlikely to have significant impact on the main galaxy at $z=0$\footnote{We emphasize the main galaxy here because early evolution in the epoch of reionization could have profound effects on the satellites.}. Hence the HN model may provide the extra regulation at high-redshift, without disrupting the expected agreement in the low-redshift Universe. The HN model is not unique in this aspect as really any physical process that suppresses star formation at low metallicity will likely have a similar impact, as long as it does not fundamentally alter subsequent galaxy evolution. However, HN may produce unique chemical signatures in low-metallicity stars that can be observed at $z=0$.

\paragraph{Ly$\alpha$ Radiation Pressure} Another physical mechanism that preferentially operates at low metallicities is Ly$\alpha$ radiation pressure \citep{Dijkstra2008,Smith2017_lya,Kimm2018,Tomaselli2021}. Energetic photons from young stars ionize their surrounding gas and Ly$\alpha$ photons are subsequently produced predominantly by recombination, but also collisional excitation. To escape the system, the Ly$\alpha$ photons must diffuse both spatially and in frequency. Due to the extreme optical depths that may be present near H~{\small II} regions, Ly$\alpha$ photons may scatter hundreds (or more) times, imparting momentum into the gas and driving outflows. However, dust can destroy Ly$\alpha$ photons by absorbing them and re-radiating at longer wavelengths. Hence, the presence of dust can significantly reduce the number of scatterings for each Ly$\alpha$ photon. For this reason, Ly$\alpha$ radiation pressure can only operate efficiently at low-metallicity \citep[e.g.][]{Kimm2018}.  

The cumulative stellar mass growth normalized by the fiducial model for our run with Ly$\alpha$ radiation pressure is shown in the second row of the middle panel of Figure~\ref{fig:reg_all}. The stellar mass growth is very similar to the previous model with HN feedback and by $z=6$, the stellar mass formed in the Lagrange (zoom-in) volume is less than half that of the fiducial model. Should we run the simulation for longer, this model will begin to approach the fiducial run as the increase in metallicity drastically weakens the impact of Ly$\alpha$ radiation pressure. 

\begin{figure}
    \centering
    \includegraphics[width=0.45\textwidth,trim={0cm 0.5cm 0cm 0cm},clip]{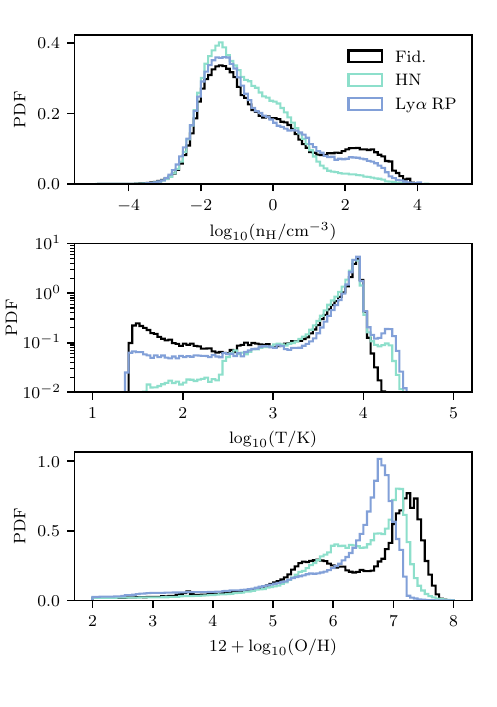}
    \caption{Mass-weighted PDFs of gas density (top), gas temperature (middle), and oxygen abundance (bottom) inside haloes in the redshift interval $6.5\leq z\leq8.5$. Including additional, strong feedback channels such as hypernovae (green) or Ly$\alpha$ radiation pressure (blue) strongly reduces the amount of dense and cold gas, while generating more hot and diffuse gas. This also leads to a suppressed stellar mass (Figure~\ref{fig:reg_all}).}
    \label{fig:rho_t_hn_lya}
\end{figure}

Ly$\alpha$ radiation pressure operates early on in the star formation cycle as ionizing photons are required to produce significant quantities of Ly$\alpha$ radiation. Furthermore, there is a strong density dependence of the effect as the optical depth is highest in dense gas. We therefore expect that the properties of the ISM is fundamentally different. 

In Figure~\ref{fig:rho_t_hn_lya}, we show the mass-weighted PDFs of gas density, temperature, and oxygen abundance for gas inside of haloes in the redshift interval $6.5\leq z\leq8.5$. We compare the fiducial model with the simulations that include HN or Ly$\alpha$ radiation pressure. This redshift interval was chosen to be between the two mergers when the galaxy is more settled to isolate effects outside of starburst periods that may be more generally representative of galaxy formation. Indeed, we find that there is a significant reduction in dense gas in the runs with HN and Ly$\alpha$ radiation pressure. This is due to two reasons. First, the stronger or additional feedback processes both preferentially operate in dense gas and this is a clear signature of the coupling of the feedback. In the case of Ly$\alpha$ the multiplicative factor for the radiation pressure increases with gas density. With regards to HN, since these only occur for the most massive stars, when they do occur, they will be the first explosions for an individual star particle. We thus empirically find that a HN is more likely to explode in gas with, for example, $n_{\rm H}>10^2\ {\rm cm^{-3}}$ than a normal SN.  Second, because fewer stars have formed in the runs with HN and Ly$\alpha$ radiation pressure, the galaxies are less chemically evolved and metal-line cooling is less efficient which slows the collapse. This is true even though HN inject significantly more O than SNe and the  gas fractions are different. PDFs of oxygen abundance are shown in the bottom panel of Figure~\ref{fig:rho_t_hn_lya}.

Differences are also apparent in the gas temperature distribution. In all three simulations, the warm phase dominates. Consistent with the reduction in high density gas seen for the HN and Ly$\alpha$ radiation pressure models, there is less cold gas in both of these simulations as well compared to the fiducial model. A key difference is that a second peak has emerged in the runs with HN and Ly$\alpha$ radiation pressure that separate the warm neutral medium from the ionized gas. When the dense gas is cleared by the simulations, the Strömgren spheres are much better resolved and more ionized gas emerges in the ISM. This gas does not cool as efficiently as in the case where the Strömgren spheres are unresolved which can help regulate star formation.  

Our results are consistent with that found in the isolated galaxy simulations of \cite{Kimm2018}. More specifically, their simulations with SN and Ly$\alpha$ radiation pressure formed fewer stars than their model without Ly$\alpha$ radiation pressure. Likewise, a key signature of the Ly$\alpha$ radiation pressure was that the dense gas clumps which tended to form bound star clusters were often destroyed in the runs with the additional feedback. However, our implementation of Ly$\alpha$ radiation pressure possibly falls on the optimistic side of what one may expect. Because it is generally too expensive to run Monte Carlo radiative transport on-the-fly in galaxy formation simulations (although c.f. \citealt{Smith2018dd} for an alternative approach), we have relied on a subgrid model for Ly$\alpha$ radiation pressure that is based on high-resolution idealized simulations. More specifically, we have adopted a uniform spherical geometry that effectively maximizes the amount of pressure that can be obtained. In gas with a turbulent geometry, Ly$\alpha$ photons make leak before they scatter enough times to have a dynamical effect on the gas. Second, the Ly$\alpha$ emissivity from recombination scales as a power-law with gas temperature such that at lower temperatures, more Ly$\alpha$ is produced. When the Strömgren spheres are not fully resolved, the temperature represents a mixed of the ionized and neutral phases and may be artificially low, which could result in an overestimate of the number of Ly$\alpha$ photons. In principle, this would also impact collisional production of Ly$\alpha$ but this is expected to be subdominant in H~{\small II} regions and is thus not included in our model. One must also keep in mind that the production of Ly$\alpha$ is related to the input number of ionizing photons. Here we have used {\small STARBURST99} \citep{Leitherer1999}; however other models, \eg {\small BPASS} \citep{Stanway2018,Eldridge2017,Eldridge2008} are known to produce more ionizing photons per stellar mass, while IMF variations can also impact the ionizing photon budget. For this reason, the numerical over-prediction in Ly$\alpha$ may in part be balanced by uncertainties in the input SED.

\subsection{Star Formation Algorithm}
Our previous model variations have concerned changes to the star particle mass and feedback. We now consider how the star formation algorithm impacts early galaxy formation. Details of the three additional models we consider are listed in Section~\ref{sec:sf_alg}, but briefly, the Jeans and Strömgren models adopt star formation efficiencies per free-fall time that scale more strongly with density compared to the fiducial model, while the density-based model employs a fixed density threshold and efficiency.

The right panel of the second row of Figure~\ref{fig:reg_all} shows the cumulative stellar mass growth as a function of time normalized by the fiducial model. The final stellar masses at $z=6$ are relatively independent of these three modifications of the star formation algorithm. Although the tested star formation algorithms are different, these results are consistent with \cite{Hopkins2018} who argue that star formation is feedback-regulated and the final stellar mass should be relatively insensitive to the details of the algorithm as long as stars form at high enough density. 

\begin{figure}
    \centering
    \includegraphics[width=0.45\textwidth]{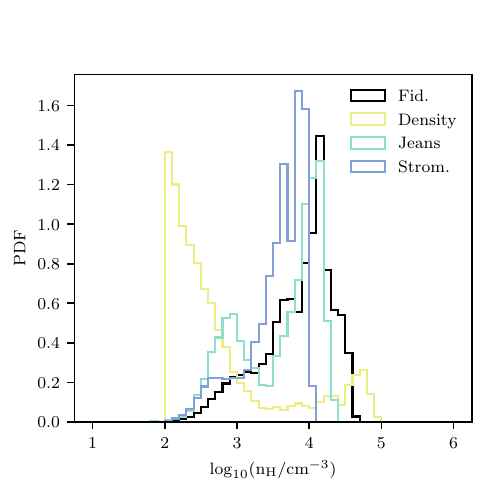}
    \caption{PDF of gas densities where stars form. Compared with our fiducial model (black), alternate star formation algorithms lead to large differences in the distribution of star-forming gas and thus the physical conditions of their H~II regions. }
    \label{fig:sf_dens_jean_strom}
\end{figure}

This result is quite interesting because both the densities of star formation and the properties of the ISM are fundamentally different between the models that adopt different star formation algorithms (see also \citealt{Hopkins2013}). In Figure~\ref{fig:sf_dens_jean_strom}, we show the PDF of the densities at which star particles form for the three alternative star formation algorithms compared to the fiducial model. The Jeans and Strömgren models have a similar shape to the fiducial model albeit biased towards slightly lower gas densities. This is an inevitability of making the efficiency scale strongly with gas density. In contrast, when applying the density based star formation model, most stars form near the density threshold that we impose.

\begin{figure}
    \centering
    \includegraphics[width=0.45\textwidth]{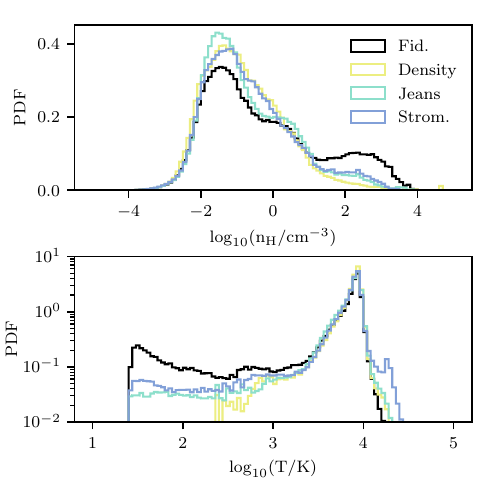}
    \caption{Mass-weighted PDFs of gas density (top) and gas temperature (bottom) inside haloes in the redshift interval $6.5\leq z\leq8.5$. Despite the small changes in final stellar mass, the ISM structure is fundamentally modified by swapping between our fiducial model (black) and other star formation algorithms (colored lines). This has direct consequences on the resulting observable emission lines of the galaxy (Section~\ref{sec:discussion:jwst}).}
    \label{fig:ism_dens_jean_strom}
\end{figure}

Changing the SF algorithm clearly impacts the properties of the ISM. In Figure~\ref{fig:ism_dens_jean_strom}, we show mass-weighted density and temperature PDFs of gas within haloes in the redshift interval $6.5\leq z\leq8.5$. The model with a density threshold for star formation has the least amount of high-density gas and almost no cold gas below \SI{500}{K} because this gas is rapidly turned into stars. Similar, but less severe deficits are also observed in the Jeans and Strömgren star formation models. These latter two models are quite similar in terms of their ISM properties, except for the fact that the ionized peak is much more prevalent in the Strömgren star formation model compared to the Jeans model (by design). Even though the stars form at much lower density in the density-based star formation model, the Strömgren spheres are still generally unresolved.  

Because the density-based star formation is the most fundamentally different to the turbulence-inspired star formation criteria, we consider three additional variations changing (a) the efficiency, (b) the efficiency and the density threshold, and finally (c) the efficiency, the density threshold, and the SN energy. The cumulative stellar mass normalized to the fiducial model is shown in the bottom left panel of Figure~\ref{fig:reg_all}. Changing the efficiency to \SI{100}{\percent} and/or increasing the density threshold by a factor of 10 has no impact on the final stellar mass. In contrast, increasing the SN energy by a factor of 5 results in significant regulation, consistent with the results of Section~\ref{sec:SN_egy}. 

\subsection{Resolution}
We next study how well our simulation converges with changes to spatial and mass resolution. The bottom middle panel of Figure~\ref{fig:reg_all} shows the cumulative stellar mass formed normalized to the fiducial model for two additional simulations with a DM particle mass that is 8$\times$ higher or lower than the fiducial model while simultaneously allowing for one more or less level of refinement in the AMR hierarchy. By $z=6$, the stellar masses are well converged; however, at earlier times there are differences. For example, the onset of star formation is delayed in the low resolution model compared to the fiducial and high-resolution simulations. This is to be expected as halo assembly is delayed with lower DM resolution. 

Moreover, for a large fraction of its evolution, the high-resolution simulation exhibits up to $10\times$ lower stellar masses. Given that all stellar masses converge by $z=6$, we conclude that our model is well converged for haloes with masses $\gtrsim \SI{e9}{\Msun}$, but that there might be more significant uncertainties at lower masses. This aligns with the findings in \cite{Rosdahl2018}. With simulations sharing many characteristics to ours, they demonstrated good resolution convergence in terms of the UV luminosity function and stellar mass-halo mass relation of $\gtrsim \SI{e9}{\Msun}$ objects (see also \citealt{Kang2024}).

\subsection{Cooling function and radiative transfer}
We consider two final variations to our fiducial model. To mitigate the computational cost of solving radiative transfer, it is typical to adopt a speed of light that is significantly larger than typical gas velocities, but significantly lower than the full value of $c$ (see \eg \citealt{Rosdahl2013}). Lowering $c$ to $c_{\rm sim}<c$ is justified as long as ionization fronts move slowly, which is the case in the higher gas densities close to galaxies, which is the focus of our current study. The situation is different in the IGM where ionization front velocities can be much faster \citep[e.g.][]{Deparis2019}. 

However, sub-ionizing radiation fronts (\eg Lyman-Werner radiation) may be artificially slowed down with a reduced speed of light, as their optical depths are much lower. We thus run an additional simulation adopting $c_{\rm sim}=0.1c$, $10\times$ higher than the fiducial model ($c_{\rm sim}=0.01c$) to verify convergence against this parameter. The bottom right panel of Figure~\ref{fig:reg_all} shows the resulting stellar mass growth that is nearly identical to the fiducial model. Any numerical lag of Lyman-Werner photon fronts do not impact the global stellar mass evolution in our simulations. This is likely because metal line cooling quickly dominates over H$_2$ cooling at metallicities $\gtrsim \SI{e-4}{\Zsun}$ (see Figure~\ref{fig:trho}) and that molecular gas is not necessary for star formation once the system is metal-enriched {\protect\citep[e.g.][]{Glover2012}}. Note that \cite{Pallottini2022} estimated that under certain high-redshift conditions, the photodissociation region fronts move slower than our assumed value of $c_{\rm sim}$.

In contrast, the exact details of metal cooling may be important for the efficiency of star formation and the initial mass function \citep[e.g.][]{Safranek-Shrader2014,Sharda2023}. To further test the robustness of our results to the choice of metal cooling, we have run a simulation where we replace the non-equilibrium ion-by-ion cooling model in {\small RAMSES-RTZ} with the default cooling in {\small RAMSES-RT}. The {\small RAMSES-RT} model adopts {\small CLOUDY} \citep{Ferland1998} tables at $T>\SI{e4}{K}$ that are interpolated over gas density and temperature, assuming scaled solar abundance patterns. At $T<\SI{e4}{K}$, the analytic cooling function from \cite{Rosen1995} is used.

The bottom right panel of Figure~\ref{fig:reg_all} demonstrates that the stellar mass growth in the simulation is reasonably independent of the chosen cooling model. These results again agree with \cite{Hopkins2018}, where various simple toy models were adopted to understand the impact of cooling on global galaxy evolution and only small changes were found.

\begin{figure}
    \centering
    \includegraphics[width=0.45\textwidth,trim={0cm 6.9cm 0cm 0cm},clip]{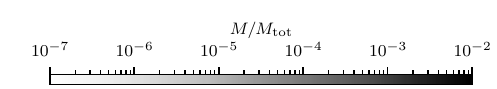}
    \includegraphics[width=0.45\textwidth]{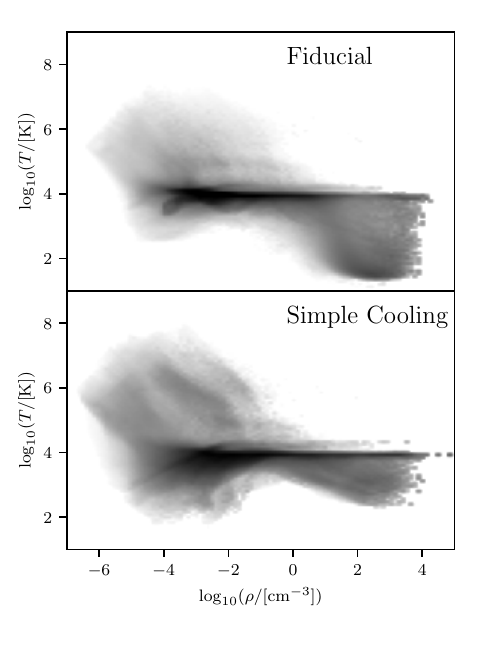}
    \caption{2D mass-weighted histograms of gas density and temperature for all gas in the Lagrange (zoom-in) volume of the main halo between $6\leq z\leq8$. Replacing the non-equilibrium cooling of RAMSES-RTZ (top) with an analytic cooling function (bottom) reduces the amount of cold, dense gas, changing the ISM structure, even if the total stellar mass of the galaxy is remains the same (Figure~\ref{fig:reg_all}). }
    \label{fig:cool_comp}
\end{figure}

Nevertheless, cooling does have a strong impact on the structure of the ISM. For example, 1D models of \cite{Kim2023} show how the ISM can lose its multi-phase nature under certain assumption regarding the cooling functions. These effects are, to some extent, subdued in 3D in the presence of turbulence, gravitational compression, and various other galaxy formation processes. To visualize these distinctions, Figure~\ref{fig:cool_comp} shows 2D mass-weighted histograms of gas density and temperature for all gas in the Lagrangian region of the main halo in the redshift interval $6\leq z\leq8$.  Despite the similarities in final stellar mass, the ISM structure is starkly different between cooling functions. Dense gas in our fiducial model becomes much colder, more rapidly above a density of $\sim\SI{1}{cm^{-3}}$ where thermal instabilities occur. In contrast, the simple cooling model begins efficient cooling at slightly lower densities, but the decrease in temperature is more gradual, consistent with expectations from the 1D calculations of \cite{Kim2023}. This clearly demonstrates that the cooling strongly impacts the thermochemistry of the ISM, even if stellar mass is converged. Excitingly, such differences would have an out-sized impact on observable properties such as collisionally excited emission lines where emissivity depends exponentially on gas temperature. We turn next to quantifying how to distinguish models based off their emission line ratios.

\section{Constraining high-redshift Galaxy formation models with ism observations} \label{sec:discussion:jwst}

Prior to JWST, constraints on early galaxy formation primary came from photometric data in the rest-frame UV from HST for hundreds of galaxies and ALMA detections of IR emission lines for a handful of objects. The inferred stellar mass-halo mass relation \citep[e.g.][]{Behroozi2013,Behroozi2019,Moster2013,Moster2018} has commonly been used at high redshift; however, there are clear systematic effects, often related to the assumed star formation history model (or prior in the case of a non-parametric model) that lead to $>\SI{1}{dex}$ uncertainty in stellar masses \citep{Narayanan2024}. The UV luminosity function represents an alternative \citep[e.g.][]{Rosdahl2018}, although the bright end is highly sensitive to assumptions on dust. Multiple simulations have been able to reproduce the UV luminosity function, often by calibration, despite fundamental differences in the underlying model \citep[e.g.][]{Gnedin2014,Rosdahl2018,Pawlik2017,Kannan2022}. With JWST, we can move to much larger samples of galaxies with photometric (in multiple bands) and spectroscopic data. For example \cite{spdrv1} showed that matching galaxy colors is significantly more difficult than UV luminosity since the latter is highly dependent on both dust and star formation history. Moreover, seeing redder with JWST compared to HST allows us to extend the baseline of available data deep into the epoch of reionization. Models that may have performed well at $z=6$ often struggle at $z>9$ \citep[e.g.][]{Finkelstein2023,Weibel2024}. 

The properties of the ISM hold a significant amount of information on the conditions of star formation in the early Universe. For simulations that match the UV luminosity function, they often struggle to reproduce certain emission line luminosity functions \citep[e.g.][]{Meyer2024}. The vast majority of large cosmological simulations of the early Universe do not model or resolve the detailed properties of the ISM \citep[e.g.][]{Gnedin2014,Kannan2022,Pawlik2017,Lovell2021} and the constraining power of the high-redshift ISM via JWST observations is thus a strong motivation for simulations that do reach the required resolutions \citep[e.g.][]{Rosdahl2018,Ma2018,Oshea2015}. Many of the variations on the fiducial model result in no significant change in stellar mass or star formation rate, but the properties of the ISM are fundamentally different which can have a large impact on observable properties. Here we demonstrate a few examples of how certain models can be ruled in or out with emission lines. 

\subsection{The O32 diagnostic}
Both [O~{\small III}]~$\lambda$5007 and [O~{\small II}]~$\lambda\lambda$3727 are readily observable with NIRSpec or a combination of NIRSpec and MIRI at $z\gtrsim9.5$ \citep[e.g.][]{Bunker2023,Castellano2024,Zavala2024}. The O32 ratio (defined as $\log_{10}$~[O~{\small III}]~$\lambda$5007/[O~{\small II}]~$\lambda\lambda$3727) is particularly interesting in the context of our work due to its sensitivity to the ionization parameter \citep[e.g.][]{Kewley2019}. Furthermore, the critical densities of both [O~{\small II}] lines are smaller than [O~{\small III}]~$\lambda$5007. In principle, star formation occurring at higher densities and with higher efficiencies would simultaneously boost the ionization parameter and be above the critical density for [O~{\small II}] emission, leading to higher O32 ratios. 

\begin{figure}
    \centering
    \includegraphics[width=0.45\textwidth,trim={0cm 0cm 0cm 1cm},clip]{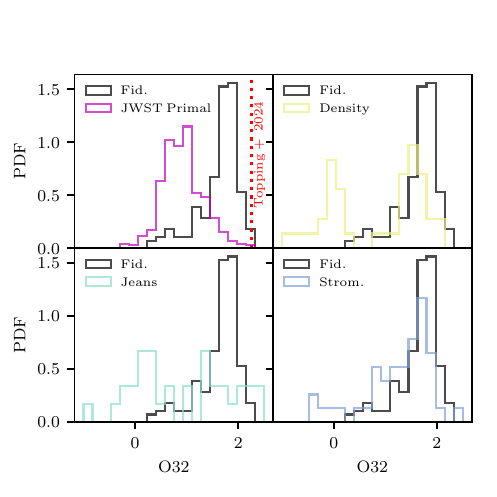}
    \includegraphics[width=0.45\textwidth,trim={0cm 0cm 0cm 1cm},clip]{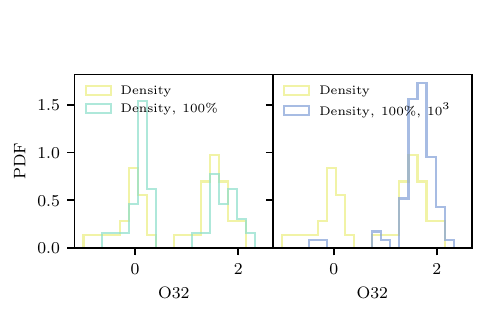}
    \caption{PDFs of O32 where we either vary the star formation algorithm (top) or vary the properties of the density-based star formation algorithm by using a higher efficiency or higher gas density threshold (bottom). Swapping algorithms generates shifts in emission line ratios that are larger than the current width of observed data (purple, from \protect\citealt{Cameron2023,Topping2024}). Even modest changes to input parameters within a given algorithm (\eg blue vs.\ green in bottom panel going from $\SI{e2}{cm^{-3}}$ to $\SI{e3}{cm^{-3}}$ in SF threshold) can have an impact on observables.}
    \label{fig:O32}
\end{figure}

In Figure~\ref{fig:O32}, we show the PDF of O32 values for all uncontaminated haloes in the Lagrange (zoom-in) volume of the main object at all redshifts with H$\alpha$ luminosities $>\SI{e39}{erg.s^{-1}}$ for the fiducial model. The luminosity cut is applied to ensure that the galaxy is undergoing an active starburst and there are on average 70 galaxies in each PDF. We contrast them with the same objects evolved with different star formation algorithms. The fiducial model (black) tends to form stars at the highest densities and this leads to the highest O32 values. In contrast, the model with the density-based star formation (yellow) generally exhibits much lower O32 values. The models with the Jeans and the Strömgren star formation algorithms form stars at slightly lower densities than the fiducial model which again shifts O32 to lower values. For comparison, we show high-redshift data from the JWST Primal database \citep{Heintz2024} in magenta and a galaxy with one of the highest O32 values ever detected at high redshift \citep{Topping2024} in red. While our one simulated galaxy is in no way fully representative of the observed galaxy population, it reaches a UV magnitude that is currently accessible by gravitational lensing in both spectroscopic and photometric surveys \citep[e.g.][]{Bezanson2022,Vanzella2023,Naidu2024_alt}. Nevertheless, we recognize that our single object and relatively few data points may limit the generalizability of our results especially considering that the H$\alpha$ luminosities and UV magnitudes of galaxies in the JWST Primal database are typically brighter than what we have simulated. However, it is clear from our numerical experiments that observational data can be used to constrain the physical inputs into numerical simulations (should we be able to simulate more massive galaxies) beyond what can be done with integrated stellar masses and UV luminosity functions. 

We emphasize that O32 PDF is highly sensitive to even minor variations in the star formation algorithm, and of parameters within a given algorithm. For example, in the bottom panel of Figure~\ref{fig:O32} we show the PDF of O32 for the density-based star formation algorithm (yellow) where we have either increased the efficiency from \SI{10}{\percent} to \SI{100}{\percent} (green) or we have increased the density threshold from \SI{e2}{cm^{-3}} to \SI{e3}{cm^{-3}} and the efficiency to \SI{100}{\percent} (blue). The shapes of the PDFs are different in all three cases. More quantitatively, we have run a two-sample KS test comparing the O32 distributions of the modified simulations to the fiducial models and found that the $p-$values are always less than 0.05. This demonstrates that the ionized gas can also probe the star formation efficiency is a similar way to the molecular gas \citep{Hopkins2013-mol,Onus2018}. Modifying the value of the density threshold, however, leads to a stronger change in O32 compared to increasing the efficiency. 

\subsection{The C43 diagnostic}
\begin{figure}
    \centering
    \includegraphics[width=0.45\textwidth,trim={0cm 0cm 0cm 1cm},clip]{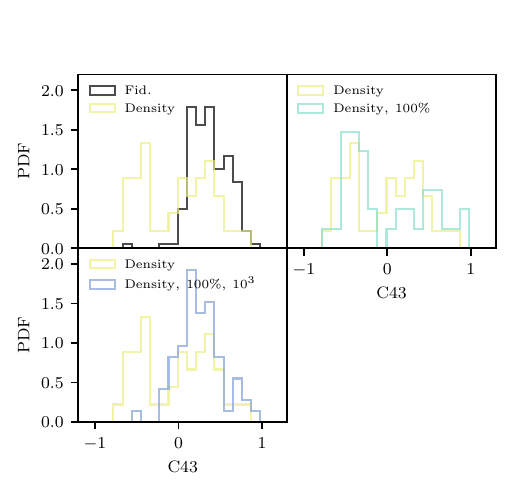}
    \caption{Same as Figure~\ref{fig:O32}, but for the C43 line ratio.}
    \label{fig:C43}
\end{figure}

Like O32, the ratio of $\log_{10}$~C~{\small IV}~$\lambda\lambda$1550/[C~{\small III}]~$\lambda\lambda$1908 is also a tracer of ionization parameter. The ionization potential of C$^{3+}$ is \SI{47.9}{eV} compared to \SI{35.1}{eV} for O$^{++}$ and hence this ratio traces higher ionization states compared to O32. Being UV lines, C43 can be observed to higher redshift than O32 with NIRSpec. Moreover, some have suggested that C43 is a tracer of LyC escape \citep{Schaerer2022}. In Figure~\ref{fig:C43} we show PDFs of C43 for the fiducial model compared with variations of the density-based star formation algorithm. While the fiducial model predicts that nearly all star-forming galaxies will have stronger C~{\small IV} emission, depending on the parameters of the density-based model, either C~{\small IV} or [C~{\small III}] can dominate. The trends are very similar to that of O32 although the spread is slightly less pronounced. KS tests again show $p-$values $<0.05$ when comparing the density-based star formation models to the fiducial model.

One possible difference between the ratios of O32 and C43 is how the elements are enriched. Oxygen comes primarily from SN while C is enriched by SN and AGB winds \citep[e.g.][]{Kobayashi2020}. For a supernova, the relevant timescale is when the shell cools and begins to re-integrate into the ISM. This timescale scales inversely with gas density\footnote{The exponent on density is $\sim-0.55$.} \citep[e.g.][]{Thornton1998,Kim2015} and is $\sim50$ times shorter for a density of $\SI{e5}{cm^{-3}}$ compared to $\SI{e2}{cm^{-3}}$. SN will on average explode at higher densities when a higher density threshold is used for star formation which will lead to the metallic ejecta from SNe being re-illuminated on a shorter timescale. If the timescale is much shorter than the lifetime of a massive star, detecting extremely metal-poor galaxies will become more difficult. In principle, the different density thresholds for star formation will lead to different timescales for this effect, but higher resolution simulations that resolve each SN remnant will be needed to confirm whether this is expected to happen. 

\subsubsection{The ${\rm [O\ III]\ 4363/5007}$ diagnostic}
While O32 probes ionization parameter and gas density, $\log_{10}{\rm [O\ III]\ 4363/5007}$ provides insight into gas temperature\footnote{Note that this only true up to gas densities of $\sim10^5\ {\rm cm^{-3}}$ (see e.g.\ \citealt{Katz2023_jwst}).}. This ratio has now been observed for numerous high-redshift galaxies \citep[e.g.][]{Katz2023_jwst,Curti2023, Laseter2024, Morishita2024}. Different feedback models can subtly change the temperature of the ISM and because of the strong sensitivity of this metric to gas temperature, it may help discriminate between models. 

\begin{figure}
    \centering
    \includegraphics[width=0.45\textwidth,trim={0cm 0cm 0cm 1cm},clip]{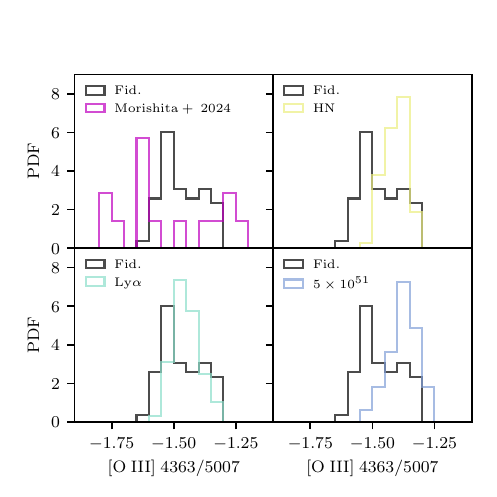}
    \caption{${\rm [O\ III]\ 4363/5007}$ ratio for our fiducial model (black). This emission line ratio is highly sensitive to the temperature structure of the ISM -- boosting the feedback energy with either HN feedback (yellow), Ly$\alpha$ radiation pressure (green), or a $5\times$ boosted SN energy systematically shifts the PDF to higher ratios and thus higher electron temperatures. We also show high-redshift observations from \protect\cite{Morishita2024} in magenta. Note that for a gas density of $500\ {\rm cm^{-3}}$, electron temperatures of $10^4$, $10^{4.3}$, and $10^{4.5}$~K correspond to ${\rm [O\ III]\ 4363/5007}$ ratios of -2.185, -1.475, and -1.217, respectively.}
    \label{fig:O33}
\end{figure}

In Figure~\ref{fig:O33} we show a PDF of ${\rm [O\ III]\ 4363/5007}$ for galaxies in the fiducial model compared with models with HN (yellow), Ly$\alpha$ radiation pressure (green), and a $5\times$ boost in the SN energy (blue). In general, the fiducial model has the lowest gas temperatures, followed by the model with Ly$\alpha$ radiation pressure. The highest temperature gas is found in the models with HN and $5\times$ boost in the SN energy. As above, KS tests indicate that these distributions are quantitatively different. There are a few reasons why these differences occur. First the models with HN and a $5\times$ boost in the SN energy simply inject more energy into the ISM and hence the gas temperature can be higher. The effect is less extreme for the Ly$\alpha$ radiation pressure model. Second, the fiducial model is more chemically evolved than the other models and the enhanced metallicity means that the ISM can cool more efficiently. Hence this metric encapsulates both metallicity and energy budget.  

For comparison, we show observational data from \cite{Morishita2024} at $z>6$ for systems where both [O~{\small III}] lines have been detected. Again, our small zoom simulations cannot be representative of the observed galaxy population, but it is clear that different feedback channels lead to systematically distinct ISM temperatures. 

\vspace{1cm}
\subsection{Discussion on Observational Diagnostics}
We have demonstrated that the observable properties of the ISM are sensitive to the detailed implementation of galaxy formation physics. If samples of simulated galaxies can be created that match the selection functions of observed high-redshift galaxies, there is an opportunity to use the observable properties to constrain the input physics and the structure of the ISM in simulations. Here we have focused primarily on emission line ratios from oxygen as they are readily observable at high-redshift and they are less sensitive to the exact details of early chemical enrichment for which there remain significant uncertainties. Nevertheless, other line ratios observable by JWST from C$^{++}$ and N$^{3+}$ probe different gas densities and may help construct a density PDF that will be even more constraining than a single line ratio. Furthermore, the exact chemical abundances are sensitive to enrichment channel and initial mass function and could thus be used to further constrain high-redshift physics. Here we only aimed to demonstrate a few examples of what is now possible. One of the key advantages of {\small RAMSES-RTZ} is that it can predict these emission line ratios self-consistently without post-processing, which opens a new avenue to understand early galaxy evolution.

\section{Conclusions}
We have introduced a new model ({\small MEGATRON}) for cosmological simulations of galaxy formation with on-the-fly multi-frequency radiative transfer and non-equilibrium primordial, molecular, and metal chemistry. Key features of the model are that the thermochemistry is designed to reproduce the conditions of the local ISM at solar metallicity (a probe of neutral gas; Figure~\ref{fig:trho}) and it is consistent with 1D photoionization models used to predict nebular emission lines from H~{\small II} regions (a probe of ionized gas; Figure~\ref{fig:line_cal}). Furthermore, star formation is based on turbulent criteria that are designed to mimic results from higher resolution simulations \citep{Federrath2012,Padoan2011}, while the adopted feedback prescription follows a model that has been shown to reproduce many of the properties of Milky Way-like galaxies at $z=0$ \citep{Agertz2013, Agertz2015, Agertz2021}. We run numerous simulations to understand how the model performs in the context of high-redshift galaxy formation and to changes in the fiducial parameters. Our work focuses on a single halo that reaches a halo mass of $\sim10^9\ {\rm M_{\odot}}$ at $z=6$ with a stellar mass up to $\sim10^7\ {\rm M_{\odot}}$, depending on model. By considering the entire history of the halo and uncontaminated neighbours within the Lagrange regions, we analyze $\sim70$ galaxies per simulation. Our primary conclusions are as follows:
\begin{enumerate}
    \item Within the context of our model, numerous parameter choices such as the mass of the star particle, the star formation efficiency per free-fall time, the timescale over which SN explode, resolution, cooling function, and star formation algorithm have little impact on the final stellar mass of the object (Figure~\ref{fig:reg_all}). Deviations may occur in the star formation history at earlier times when the halo is less massive, but once the halo reaches a mass of $\sim \SI{e9}{\Msun}$, the stellar mass converges. Thus, in high-redshift conditions with significant inflows and mergers, the spatial and temporal clustering of stellar feedback, star formation algorithm, and detailed cooling physics are secondary parameters for star formation regulation, even if they may induce large shifts in gas densities at which stars form and SN explode. 
    
    \item The total feedback energy budget is a key parameter that controls the final stellar mass of the halo. Either arbitrarily increasing the SN energy, or accounting for additional feedback channels (e.g.\ hypernovae, Ly$\alpha$ radiation pressure) both suppress the final stellar mass by reducing the dense, cold gas content of the halo (Figure~\ref{fig:gas_fracs} and~\ref{fig:rho_t_hn_lya}). Impulsivity also matters, i.e. we find that the HN are more effective at regulating star formation as they inject more energy in a single event compared to simply increasing the energy budget of each SN by a smaller amount. The effect is also non-linear in that a threshold amount of energy must be injected before significant differences in stellar mass are seen. For our simulations, this seems to occur at $5\times10^{51}$~erg per SN.

    \item Even if most model variations and parameter choices yield modest changes to the integrated stellar mass, the structure of the ISM is very sensitive to these choices. In particular, the densities of gas surrounding young stars, and the ISM density and temperature PDFs can vary strongly with the choice of star formation algorithm (Figure~\ref{fig:ism_dens_jean_strom}) or the adopted cooling function (Figure~\ref{fig:cool_comp}). These differences modify the distribution function of strong emission line ratios that are particularly sensitive to density and ionization parameter (\eg O32, C43; Figure~\ref{fig:O32}, \ref{fig:C43}) and temperature ([O~{\small III}]~4363/5007; Figure~\ref{fig:O33}). Such line ratios are now observable to $z\gtrsim10$ with JWST. 
\end{enumerate}

Taken together, our results bridge a necessary gap between high-redshift numerics and the current observational progress following the advent of JWST. Moving forward, ISM observations will represent a new and powerful tool to constrain the detailed physics of galaxy formation simulations, offering a complementary angle to obtain a complete understanding of the ISM and feedback across all cosmic epochs. This work also showcases the power of accounting for non-equilibrium chemistry and radiative transfer in galaxy simulations, paving the way for future simulations of the early Universe using {\small RAMSES-RTZ} that can directly be compared to JWST photometric and spectroscopic data.

\vspace{1cm}
\section*{Acknowledgments}
This work made extensive use of the dp265 and dp016 projects on the DiRAC ecosystem. HK is particularly thankful to Christopher Mountford and Alastair Basden for support on DIaL3 and Cosma8, respectively. HK and MR thank Jonathan Patterson for support with Glamdring. HK thanks Julien Devriendt, Adrianne Slyz, Yohan Dubois, Andrey Kravtsov, and Mark Krumholz for enlightening discussions. The authors thank the referee for their constructive comments.

MR is supported by the Beecroft Fellowship funded by Adrian Beecroft. OA and CC acknowledge support from the Knut and Alice Wallenberg Foundation, the Swedish Research Council (grant 2019-04659), and the Swedish National Space Agency (SNSA Dnr 2023-00164). TK was supported by the National Research Foundation of Korea (NRF) grant funded by the Korea government (No. 2020R1C1C1007079 and No. 2022R1A6A1A03053472). This work used the DiRAC@Durham facility managed by the Institute for Computational Cosmology on behalf of the STFC DiRAC HPC Facility (\url{www.dirac.ac.uk}). The equipment was funded by BEIS capital funding via STFC capital grants {\small ST/P002293/1}, {\small ST/R002371/1} and {\small ST/S002502/1}, Durham University and STFC operations grant {\small ST/R000832/1}. This work also used the DiRAC Data Intensive service at Leicester, operated by the University of Leicester IT Services, which forms part of the STFC DiRAC HPC Facility. The equipment was funded by BEIS capital funding via STFC capital grants {\small ST/K000373/1} and {\small ST/R002363/1} and STFC DiRAC Operations grant {\small ST/R001014/1}. DiRAC is part of the National e-Infrastructure.

We thank the developers and maintainers of \textsc{pynbody} (\citealt{Pontzen2013,Pontzen2022}), \textsc{NumPy} (\citealt{vanderWalt2011, Harris2020}), \textsc{SciPy} (\citealt{Virtanen2020}), \textsc{jupyter} (\citealt{RaganKelley2014}), \textsc{matplotlib} (\citealt{Hunter2007}), the Astrophysics Data Service and the arXiv preprint repository for providing open-source software and services that were used extensively in this work.

The main roles of the authors were, using the CRediT (Contribution Roles Taxonomy) system\footnote{\url{https://authorservices.wiley.com/author-resources/Journal-Authors/open-access/credit.html}}:
\textbf{Harley Katz: Conceptualization; Data curation; Formal analysis; Investigation; Writing – original draft.} 
\textbf{Martin P. Rey: Conceptualization; Writing – original draft.} 
\textbf{Corentin Cadiou: Conceptualization; Writing – original draft.} 
\textbf{Taysun Kimm: Conceptualization; Investigation; Writing – original draft.} 
\textbf{Oscar Agertz: Conceptualization; Investigation; Writing – original draft.} 

\bibliographystyle{mn2e}
\bibliography{example}

\end{document}